\documentclass[11pt,twoside,letterpaper]{article} 
\usepackage{times,fancyhdr}
\usepackage[dvips]{graphicx}
\usepackage{bm}
  
\makeatletter
\def\cleardoublepage{\clearpage\if@twoside \ifodd\c@page\else%
    \hbox{}%
    \thispagestyle{empty}%
    \newpage%
    \if@twocolumn\hbox{}\newpage\fi\fi\fi}
\makeatother
\renewcommand{\thesection}{\arabic{section}.}
\renewcommand{\thesubsection}{\thesection\arabic{subsection}.}

\def\figurename{Figure}
\makeatletter
\renewcommand{\fnum@figure}[1]{\figurename~\thefigure.}
\makeatother
\def\tablename{Table}
\makeatletter
\renewcommand{\fnum@table}[1]{\tablename~\thetable.}
\makeatother
\begin{document}
\title{
{\begin{flushleft}
\vskip 0.45in
{\normalsize\bfseries\textit{Chapter~1}}
\end{flushleft}
\vskip 0.45in
%
%
%
%
\bfseries\scshape From nuclei to nuclear pasta}}
\author{\bfseries\itshape C.O. Dorso$^1$, P.A. Gim\'enez Molinelli $^1$ and J.A. L\'opez$^2$\thanks{E-mail jorgelopez@utep.edu}\\
${}^1$Depto. de F\'isica, FCEN, Universidad de Buenos Aires, Buenos Aires, Argentina\\
${}^2$University of Texas at El Paso, El Paso, Texas, USA}

\date{}
\maketitle
\thispagestyle{empty}
\setcounter{page}{1}
\thispagestyle{fancy}
\fancyhead{}
\fancyhead[L]{In: Neutron Star Crust \\
Editors: C.A. Bertulani and J. Piekarewicz, pp. {\thepage-\pageref{lastpage-01}}} 
\fancyhead[R]{ISBN 0000000000  \\
\copyright~2012 Nova Science Publishers, Inc.}
\fancyfoot{}
\renewcommand{\headrulewidth}{0pt}
\vspace{2in}

%
\pagestyle{fancy} \fancyhead{} \fancyhead[EC]{Dorso,
G\'imenez-Molinelli and L\'opez} \fancyhead[EL,OR]{\thepage}
\fancyhead[OC]{From nuclei to nuclear pasta} \fancyfoot{}
\renewcommand\headrulewidth{0.5pt}
%
\section{Introduction}
Neutron stars are created as a final product of the death of a
massive star.  Stars die when the internal thermonuclear fusion
can no longer balance the gravitational compression.  For such
massive stars a supernova shock ejects most of the mass of the
star leaving behind a dense core.  Since during the collapse a big
part of electrons and protons turn into neutrons and escaping
neutrinos through electron capture, the residual mass tends to
have an excess of neutrons over protons, thus justifying the name
of a neutron star.

The mass of a neutron star ranges between 1 and 3 solar masses,
its radius is of the order of a $10^{-5}$ solar radius or about 10
km, which gives it an average density of $~10^{15}$ $g/cm^3$ or
about 20 times that of normal nuclei. Energy considerations
indicate~\cite{1,4} that the cores of neutron stars are topped by
a crust of about a kilometer thick where the $\beta$
decayed-produced neutrons form neutron-rich nuclear matter
immersed in a sea of electrons.

The density of such crust ranges from normal nuclear density
($\approx 3\times10^{14}$ $g/cm^3$, or $\approx \rho_0=0.15$
nucleons/fm${}^3$) at a depth of $\approx 1 \ km$ to the neutron
drip density ($\approx 4\times10^{11}$ $g/cm^3$) at $\approx 1/2 \
km$, to, finally, an even lighter mix of neutron-rich nuclei also
embedded in a sea of electrons with densities decreasing down to
practically zero in the neutron star envelope. The study of the
structure of such crust of nuclear matter at subcritical
densities, is the purpose of the present work.

It has been shown under different
approximations~\cite{1,4,2,3,5,6,7,8,9} that low-density nuclear
matter can exist in the form of non-uniform structures which have
been known as ``nuclear pastas''. Such effect appears to be due to
the interplay between attractive-repulsive nuclear and Coulomb
forces. As the density, temperature and proton fraction vary, the
stable nuclear shape changes from the uniform phase of the core to
configurations with voids filled with a gaseous phase, to
``lasagna-like'' layers of nuclear matter and gas, to
``spaghetti-like'' rods of matter embedded in a nuclear gas to
``polpetta-like'' (meatball) clumps to a practically uniformly
dissolved gaseous phase~\cite{4}. This crust structure is a
determining factor in several astronomical processes such as
star-quakes, supernova explosions, pulsar frequency and neutrino
opacity in supernova matter.

Different models have been used to investigate such low-density
phase of nuclear matter \cite{1,4,2,3,5,6,7,8,9}. In summary,
static models (e.g. liquid-drop model, Thomas-Fermi, Hartree-Fock)
find varying structures in nuclear systems using solely energy
considerations, i.e. by a balance between the different components
of the nuclear and Coulomb energies. Dynamical models, such as
quantum molecular dynamics~\cite{horo_lambda,gw-2002}, predict the
formation of the pasta phases by dynamical means.  For complete
descriptions of some of these techniques the reader is directed to
other chapters of this volume.

Unfortunately, the predictions obtained by these different
techniques are somewhat model dependent and a direct comparison
between models is not entirely feasible.  Such is the case of the
effect that the Coulomb interaction between the embedding electron
sea and the nuclear charges has on the structure of the nuclear
matter. Initially overlooked, such interaction is now known to
produce an screening effect that aids the stability of the overall
nuclear system while modifying the structure.  This interaction,
however, can only be included under different
approximations~\cite{9} depending on the model used.

To provide a practitioner's introduction to the study of the
nuclear pasta, this article presents a brief review of the
evolution of the methods used to study nucleon dynamics in
Section~\ref{nd} This is followed with a more complete description
of the extension of classical molecular dynamics to infinite
systems in Section~\ref{md} The different techniques used to
characterize the nuclear structure, such as cluster recognition
algorithms, radial correlation functions and the Minkowski
functionals are presented in Section~\ref{ssc} We reach full
functionality of the extended $CMD$ when the Coulomb interaction
is introduced through two common methodologies and its effect on
the structure is presented in Section~\ref{cp} We close the
article with some final remarks in Section~\ref{fr}

\section{Nucleon dynamics}\label{nd}
To study the nuclear structure of stellar crusts is necessary
understand the behavior of nucleons, i.e. protons and neutrons, at
densities and temperatures as those encountered in the outer
layers of stars.  This knowledge comes from decades of study of
nucleon dynamics through the use of computational models that have
evolved from simple to complex.  In this section we review this
evolution presenting a synopsis of existing models and culminating
with the model used in this work, namely classical molecular
dynamics ($CMD$).

\subsection{Evolution of models}
Several approaches have been used to simulate the behavior of
nuclear matter, especially in reactions. The initial statistical
studies of the 1980's, which lacked fragment-fragment interactions
and after-breakup fragment dynamics~\cite{Bon96}, were followed in
the early 1990s by more robust studies that incorporated these
effects.  All these approaches, however, were based on idealized
geometries and missed important shape fluctuation and
reaction-kinematic effects.  On the transport-theory side, models
were developed in the late 1990s based on classical, semiclassical
and quantum-models; because of their use in the study of stellar
nuclear systems nowadays, a side-by-side comparison of these
transport models is in order.

Starting with the semiclassical, a class of models known
generically as $BUU$ is based on the Vlasov-Nordheim
equation~\cite{Nor28} also known as the
Boltzmann-Uehling-Uhlenbeck~\cite{Ueh33}.  These models
numerically track the time evolution of the Wigner function,
$f(\vec{r},\vec{p})$, under a mean potential $U(\vec{r})$ to
obtain a description of the probability of finding a particle at a
point in phase space. The semiclassical interpretation constitutes
$BUU$'s main advantage; the disadvantages come from the limitation
of using only a mean field which does not lead to cluster
formation. To produce fragments, fluctuations must be added by
hand, and more add-ons are necessary for secondary decays and
other more realistic features.

On the quantum side, the molecular dynamics models, known as
$QMD$, generically speaking, solve the equations of motion of
nucleon wavepackets moving within mean fields (derived from Skyrme
potential energy density functional).  The method allows the
imposition of a Pauli-like blocking mechanism, use of
isospin-dependent nucleon-nucleon cross sections, momentum
dependence interactions and other variations to satisfy the
operator's tastes.  The main advantage is that $QMD$ is capable of
producing fragments, but at the cost of a poor description of
cluster properties and the need of sequential decay codes external
to $BUU$ to "cut" the fragments and de-excite them \cite{Pol05};
normally clusters are constructed by a coalescence model based on
distances and relative momenta of pairs of nucleons. Variations of
these models applied to stellar crusts can be found elsewhere in
this volume.

Of particular importance to this article is the classical
molecular dynamics ($CMD$) model, brainchild of the Urbana group
\cite{Pan90} and designed to reproduce the predictions of the
Vlasov-Nordheim equation while providing a more complete
description of heavy ion reactions.  As it will be described in
more detail in the next section, the model is based on the
Pandharipande potential which provides the ``nuclear'' interaction
through a combination of Yukawa potentials selected to correspond
to infinite nuclear matter with proper equilibrium density, energy
per particle and compressibility.

Problems common to both $BUU$ and $QMD$ are the failure to produce
appropriate number of clusters, and the use of hidden adjustable
parameters, such as the width of wavepackets, number of test
particles, modifications of mean fields, effective masses and
cross sections.  These problems are not present in the classical
molecular dynamics model which, without any adjustable parameters,
hidden or not, is able to describe the dynamics of the reaction in
space from beginning to end and with proper energy, space and time
units.  It intrinsically includes all particle correlations at all
levels, i.e. 2-body, 3-body, etc. and can describe nuclear systems
ranging from highly correlated cold nuclei (such as two
approaching heavy ions), to hot and dense nuclear matter (nuclei
fused into an excited blob), to phase transitions (fragment and
light particle production), to hydrodynamics flow (after-breakup
expansion) and secondary decays (nucleon and light particle
emission).

The only apparent disadvantage of the $CMD$ is the lack of quantum
effects, such as the Pauli blocking, which at medium excitation
energies stops the method from describing nuclear structure
correctly. Fortunately, in collisions, the large energy deposition
opens widely the phase space available for nucleons and renders
Pauli blocking practically obsolete~\cite{Lop00}, while in stellar
environments, at extremely low energies and with frozen-like
structures, momentum-transferring collisions cease to be an
important factor in deciding the stable configuration of the
nuclear matter. Independent of that, the role of quantum effects
in two body collisions is guaranteed to be included by the
effectiveness of the potential in reproducing the proper cross
sections, furthermore, an alternate fix is the use of momentum
dependent potentials (as introduced by Dorso and
Randrup~\cite{dor-ran}) when needed.

Although no theories can yet claim to be a perfect description of
nuclear matter, all approaches have their advantages and drawbacks
and, if anything can be said about them, is that they appear to be
complementary to each other.

\subsection{Classical Molecular Dynamics}\label{md}

Out of the techniques used to describe nuclear matter, only
classical molecular dynamics can describe all limiting behavior
such as non-equilibrium dynamics, hydrodynamic flow, as well as
changes of phase.  The method was introduced decades ago to study
restrictive aspects of nuclear reactions~\cite{wilets,panos}, but
was refined for realistic heavy-ion collisions and infinite-matter
properties by Pandharipande and coworkers~\cite{pandha}.

In a nutshell, the classical molecular dynamics method considers
the participating nucleons as classical particles interacting
through a two-body potential.  By solving the coupled equations of
motion of the many-body system numerically, the $CMD$ technique
can approximate the time evolution of hundreds or thousands of
interacting nucleons to study either reactions or the structure of
infinite nuclear matter from a microscopic point of view.

A key ingredient of the $CMD$ technique is the interaction
potential.  In the past, initial studies were based on argon-like
$6$-$12$ potentials~\cite{friedman}, followed by skyrme-type
potentials~\cite{lubeck}, until more phenomenologically correct
interactions were developed by Pandharipande~\cite{pandha}.  The
resulting two-body potentials were crafted as to reproduce the
empirical energy and density of nuclear matter, as well as
realistic effective scattering cross sections.

In general terms, the $MD$ investigations of confined nuclear
matter consist of placing ``nucleons'' in a container at a desired
density and endowed with a distribution of velocities as to mimic
a maxwellian distribution of energy.  After following the
trajectories of motion until thermal equilibrium is achieved, the
system is force-heated or cooled to reach a desired temperature.
Recording the position and velocities of each nucleon allows then
to use fragment-recognition algorithms to identify clusters.  The
same procedure can be implemented to study reaction between two
colliding ``nuclei'', as well as for expanding systems.

For the case of interest, i.e. for infinite systems composed of a
given ratio of protons to neutrons, a similar procedure can be
implemented using molecular dynamics in containers under periodic
boundary conditions to simulate infinite systems.

\subsubsection{The potential} The molecular dynamics model used in
the present study was first introduced in 1999~\cite{14a}, and
ever since has been very fruitful in nuclear studies of, among
other phenomena, neck fragmentation \cite{Che02}, phase
transitions \cite{16a,Bar07}, and isoscaling
\cite{8a,9a,10a,11a,12a,13a,Dor11}; all without any adjustable
parameters. Readers are directed to these references for further
details on the model; here only a brief synopsis will be
presented.

The nuclear interaction potentials~\cite{Pan90} guarantee that the
system will have nuclear-like properties such as proper binding
energy, radii and nucleon-nucleon cross sections.  The explicit
expressions of the potential is:
\begin{eqnarray*}
V_{np}(r) &=&V_{r}\left[ exp(-\mu _{r}r)/{r}-exp(-\mu
_{r}r_{c})/{r_{c}}
\right] \\
& &\ \mbox{}-V_{a}\left[ exp(-\mu _{a}r)/{r}-exp(-\mu
_{a}r_{a})/{r_{a}}
\right] \\
V_{NN}(r)&=&V_{0}\left[ exp(-\mu _{0}r)/{r}-exp(-\mu _{0}r_{c})/{
r_{c}}\right] \ . \label{2BP}
\end{eqnarray*}
The Pandharipande potentials differentiate between different types
of nucleons: $V_{np}$ is the potential between a neutron and a
proton and it is attractive at large distances and repulsive at
small ones, and $V_{NN}$ is the interaction between identical
nucleons and it is purely repulsive; notice that no bound state of
identical nucleons can exist.

In terms of parameters, the cutoff radius used is $r_c=5.4$ $fm$
after which the potentials are set to zero, i.e.
$V_{NN}(r>r_c)=V_{np}(r>r_c)=0$. Two different sets of parameters
for the Yukawa potentials $\mu_r$, $\mu_a$ and $\mu_0$, were
developed~\cite{Pan90} and correspond to infinite-nuclear matter
systems with an equilibrium density of $\rho_0=0.016 \ fm^{-3}$, a
binding energy $E(\rho_0)=16$ MeV/nucleon and compressibility of
about $250 \ MeV$ ("Medium") or $535 \ MeV$ ("Stiff"). It should
be remarked that, at a difference from potentials used by other
models~\cite{horo_lambda}, the potential used here has a hard core
not present in other approximations.

\subsubsection{Ground state nuclei} 
\begin{figure}[h]
\begin{center}
\includegraphics[width=4.0in]{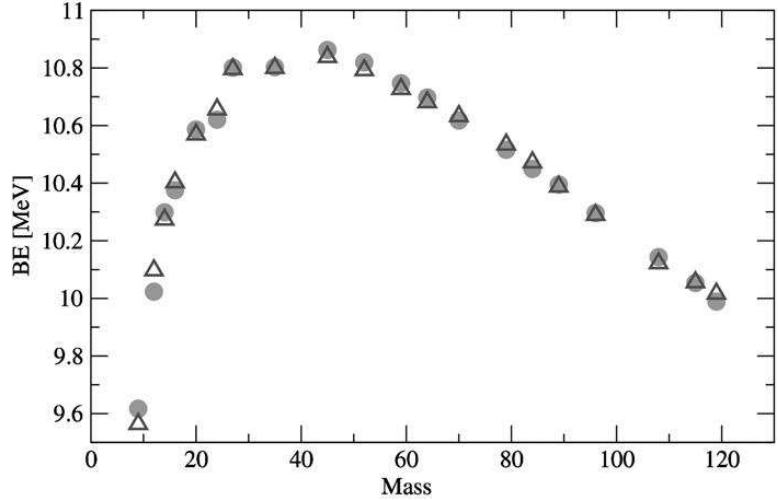} \caption{Binding
energies of ground-state nuclei obtained with mass formula fit
(triangles) for the Stiff model with the corresponding ground
states calculated using $CMD$ (circles).}\label{BE-A}
\end{center}
\end{figure}

Although the $T=0$ state of
this classical nuclear matter is a simple cubic solid, nuclear
systems can be mimicked by adding enough kinetic energy to the
nucleons. To study nuclei, for instance, liquid-like spherical
drops with the right number of protons and neutrons are
constructed confined in a steep spherical potential and then
brought to the ``ground'' state by cooling them slowly from a
rather high temperature until they reach a self-contained state.
Removing then the confining potential the system is further cooled
down until a reasonable binding energy is attained, the remaining
kinetic energy of the nucleons helps to resemble the Fermi motion.
Figure~\ref{BE-A} shows the binding energies of ``ground-state
nuclei'' obtained with the mass formula and with $CMD$;
see~\cite{Dor11} for details.

\subsubsection{Collisions} With respect to collisions, these
potentials are known to reproduce nucleon-nucleon cross sections
from low to intermediate energies~\cite{Pan90} and it has been
used extensively in studying heavy ion collisions (see
e.g.~\cite{Che02,Bar07}). For such reactions, two ``nuclei'' are
boosted against each other at a desired energy. From collision to
collision, the projectile and target are rotated with respect to
each other at random values of the Euler angles. The evolution of
the system is followed using a velocity-Verlet algorithm with
energy conservation better than 0.01\%.  At any point in time, the
nucleon information, position and momenta, can be turned into
fragment information by identifying the clusters and free
particles; several such cluster recognition algorithms have been
developed by our collaborator, C.O. Dorso, and they are well
described in the literature~\cite{Dor95,Str97}.

\begin{figure}[h]
\begin{center}
\includegraphics[width=5.in]{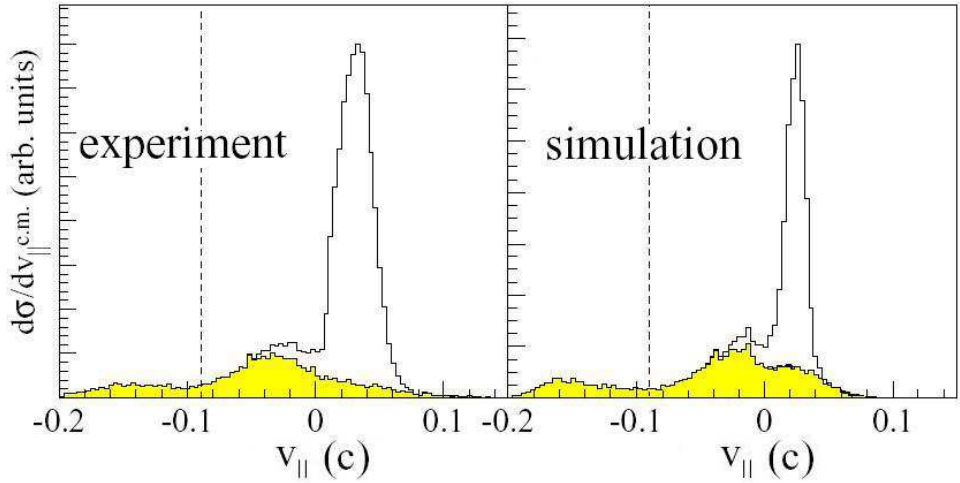}
\caption{Experimental and simulated parallel velocity
distributions for ${}^{58}Ni+C$ collisions.}\label{Cherno}
\end{center}
\end{figure}

The method yields mass multiplicities, momenta, excitation
energies, secondary decay yields, etc. comparable to experimental
data~\cite{Bel96,Che02}. Figure~\ref{Cherno}, for instance, shows
experimental and simulated parallel velocity distributions for
particles obtained from mid-peripheral and peripheral
${}^{58}Ni+C$ collisions performed at the Coupled Tandem and
Super-Conducting Cyclotron accelerators of AECL at Chalk
River~\cite{Che02}.

\subsubsection{Thermostatic properties of nuclear matter} To study
thermal properties of static nuclear matter, be these drops or
infinite systems, nucleons are positioned at random, but with a
selected density, in a container and "heated". After
equilibration, the system can then be used to extract macroscopic
variables. Repeating these simulations for a wide range of density
and temperature values, information about the energy per nucleon,
$\epsilon(\rho,T)$, can be obtained and used to construct
analytical fits in the spirit of those pioneered by Bertsch,
Siemens and Kapusta~\cite{Ber84, Kap84, Lop84, Lop00}; these fits
in turn can be used to derive other thermodynamical variables such
as pressure, etc.

\begin{figure}[h]
\begin{center}
\includegraphics[width=5.5in]{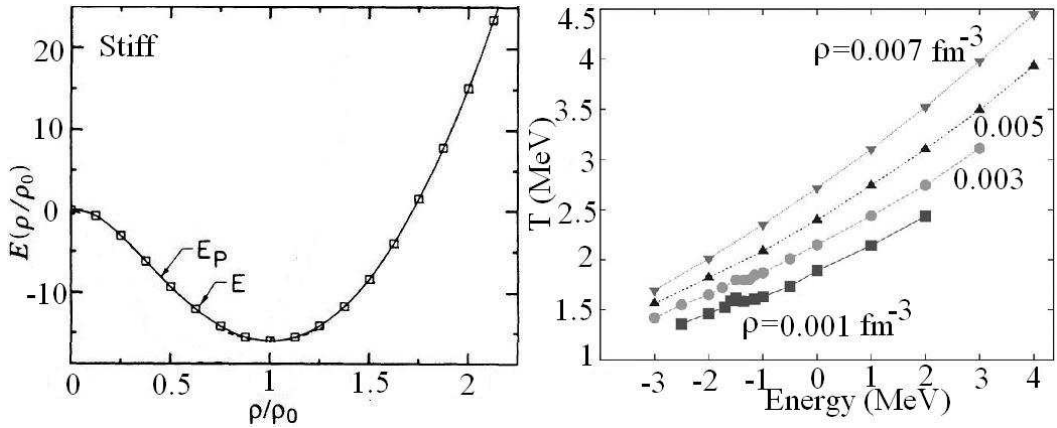}
\end{center}
\caption{Left panel: Energy per particle of stiff cold matter
calculated with $CMD$ and a polynomial fit.  Right panel: Caloric
curve of a system equilibrated at four different densities as
calculated by $CMD$.}\label{pandhador}
\end{figure}

The left panel of Figure~\ref{pandhador} shows the results of the
method as applied by Pandharipande and coworkers~\cite{Pan90} for
stiff cold infinite nuclear matter with the squares representing
$CMD$ calculations ($E$) and the line a polynomial fit ($E_p$).
For finite systems, the right panel of the same figure
demonstrates the feasibility of using $CMD$ to study thermal
properties such as the caloric curve (i.e. the temperature -
excitation energy relationship) for a system of 80 nucleons
equilibrated at four different densities; see~\cite{Dor11} for
complete details.

After having reviewed the different characteristics of $CMD$ as
well as several of its applications, we now turn to its use in the
study of nuclear structures in infinite media.

\section{Structure of stellar crusts} \label{ssc}
To study infinite systems, such as stellar crusts, $CMD$ is aplied
to systems confined in a cell surrounded with replicas to avoid
finite-size effects. The systems studied were endowed with a fixed
proton-to-neutron ratios of either $x=Z/A=0.3$ or $0.5$ and, to
always maintain a sizable proton number, this was kept at a
constant $1000$ while increasing the total number of nucleons to
$A=2000$ and $3333$ in the two different values of $x$,
respectively. As a consequence of the variation of the total
nucleon number, the size of the box was adjusted as to achieve the
desired densities; these varied between $\rho=0.01 \ fm^{-3}$ (or
$\approx \rho_0/15$) all the way to almost $\rho_0$.

The systems constructed are confined systems placed in a closed
cubical container at a fixed density and proton ratio $x$ and
allowed to equilibrate at a low temperature. The excitation energy
is added to such system by scaling the momenta of the particles.
The trajectories of individual nucleons are then obtained using an
standard Verlet algorithm with an energy conservation of
$\mathcal{O}$($0.01\%)$.

The final configurations of the system were obtained at low
temperature of $T=0.1 \ MeV$ by means of the isothermal molecular
dynamics prescribed by the Andersen thermostat
procedure~\cite{andersen}.  Our technique gradually cools the
system in small temperature steps (say $0.02 \ MeV$) while
reaching thermal equilibrium at every step; although a temperature
of $0.1 \ MeV$ is rather large for stellar crusts, in terms of the
dynamics of nucleons, it practically corresponds to a frozen
state.

\subsection{Temperature and nuclear matter structure}
\begin{figure}[h]
\begin{center}$
\begin{array}{cc}
\includegraphics[width=2.5in]{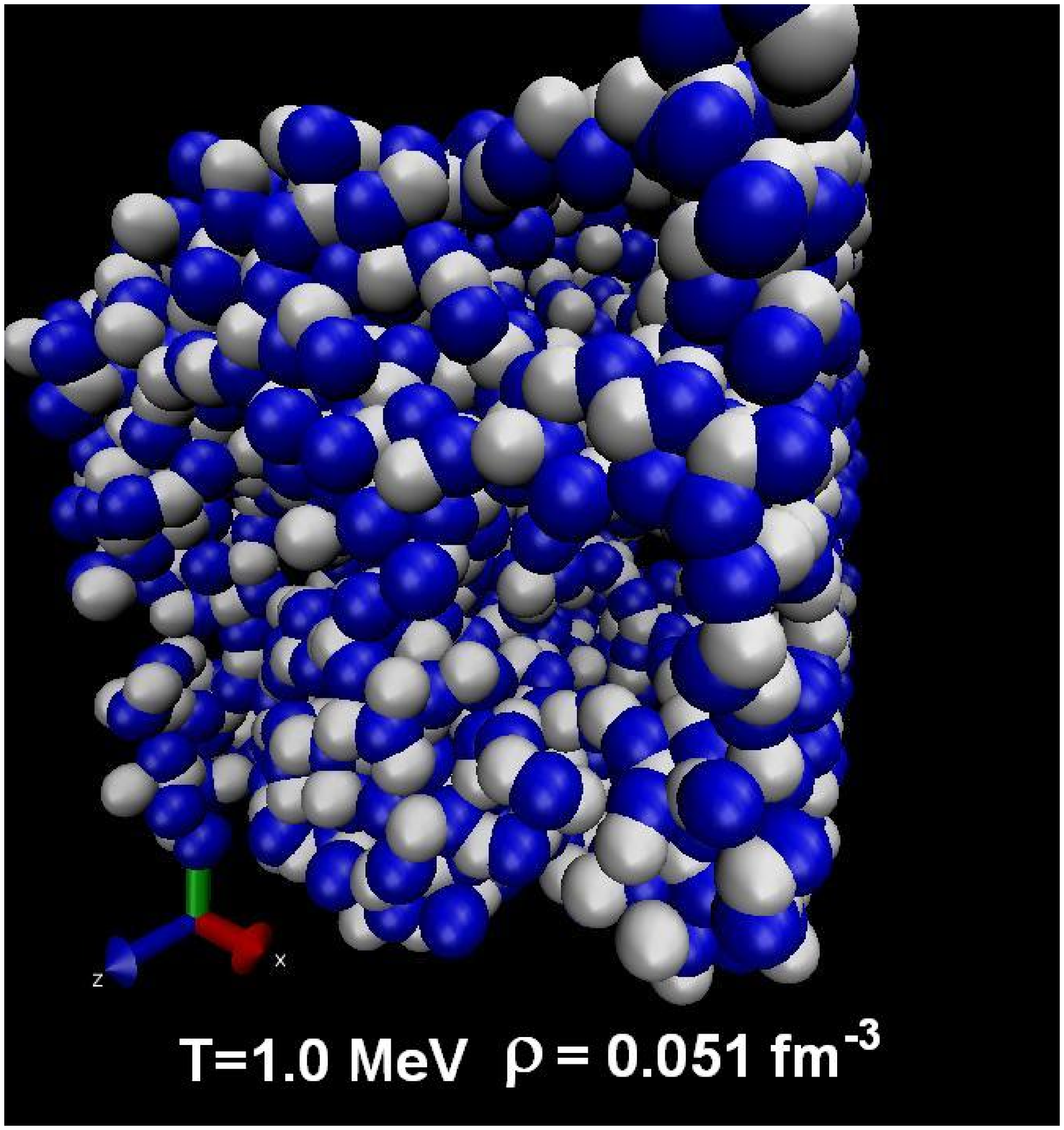} &
\includegraphics[width=2.5in]{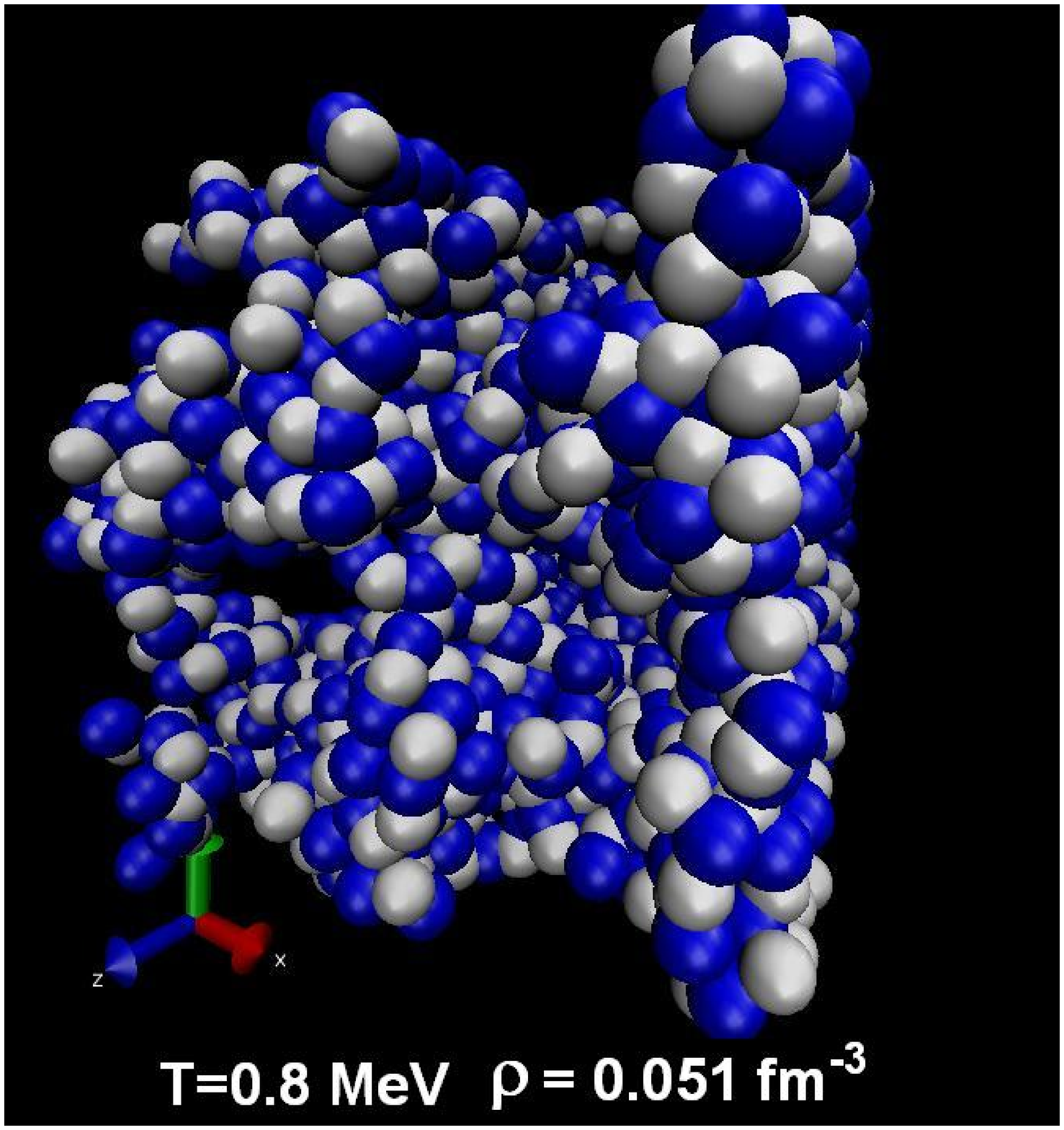} \\ \includegraphics[width=2.5in]{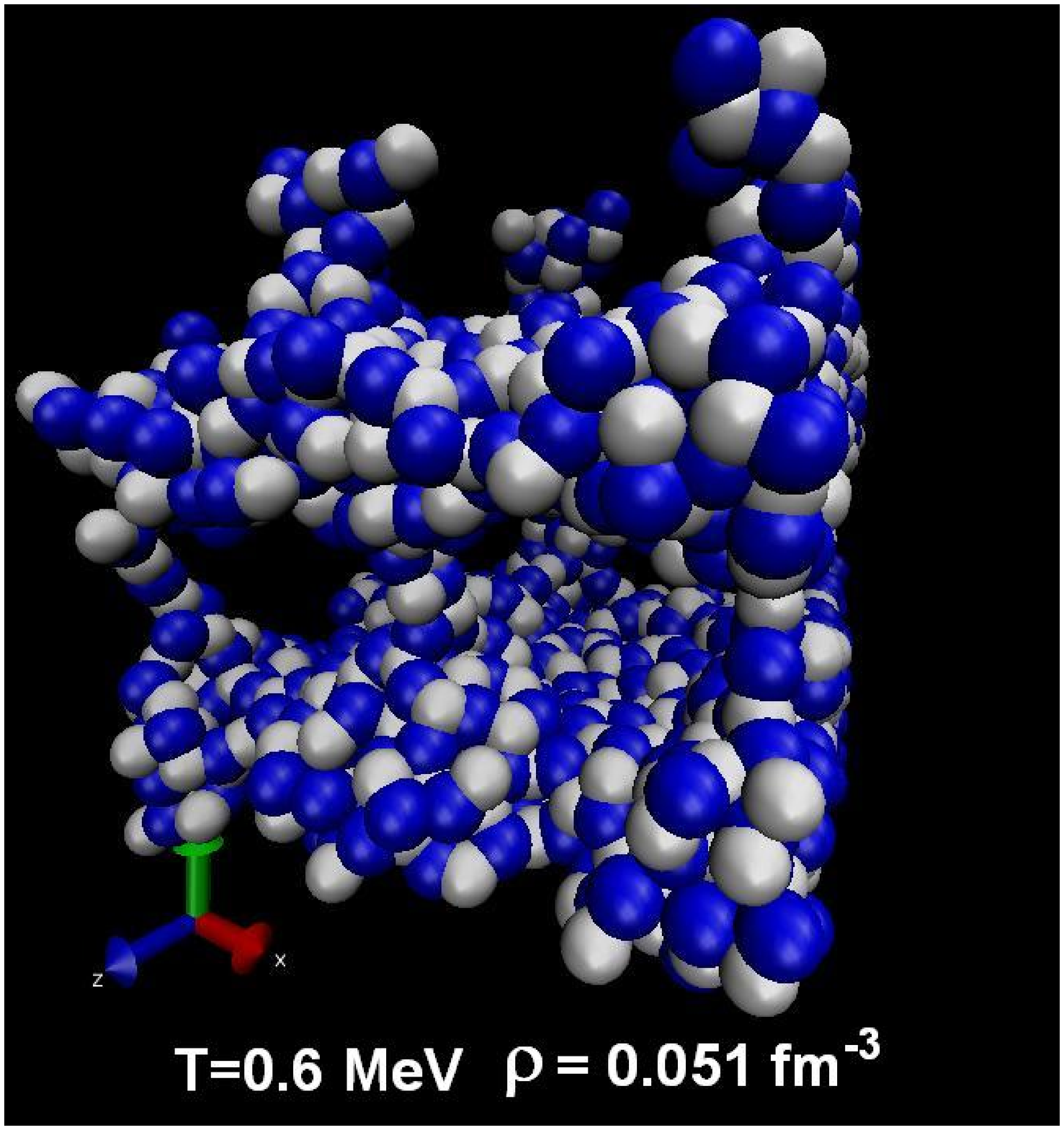} &
\includegraphics[width=2.5in]{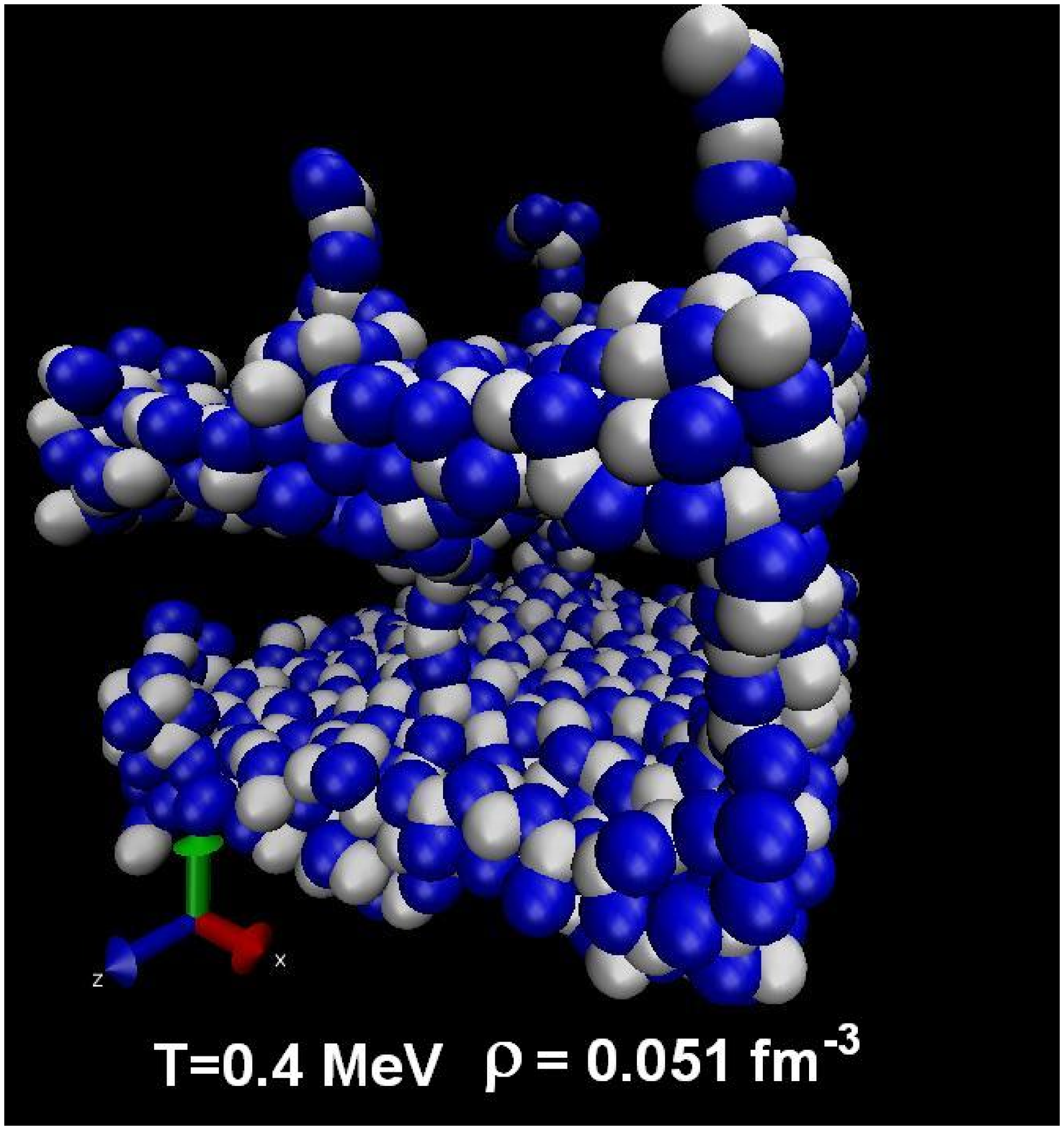} \\ \includegraphics[width=2.5in]{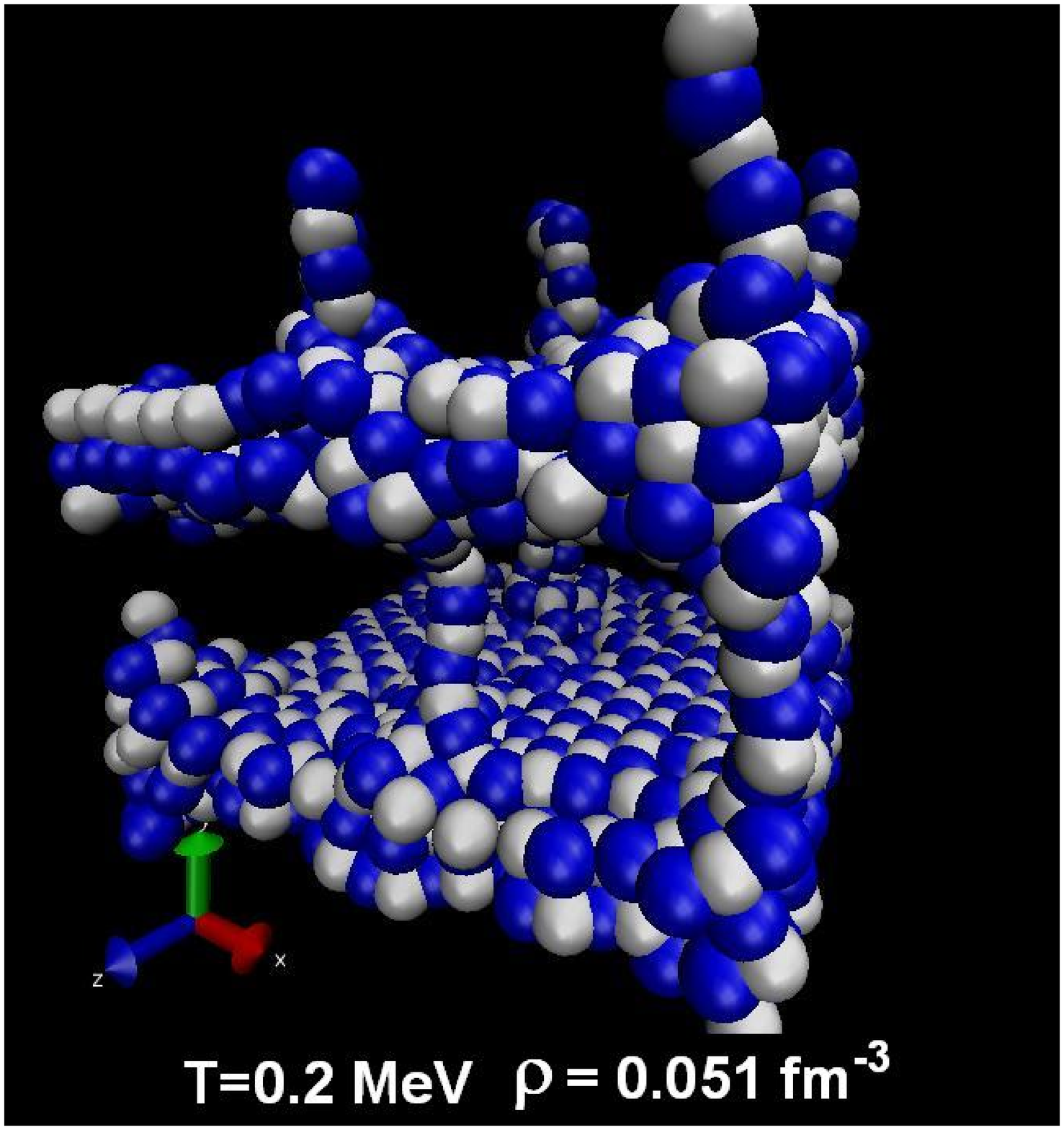} &
\includegraphics[width=2.5in]{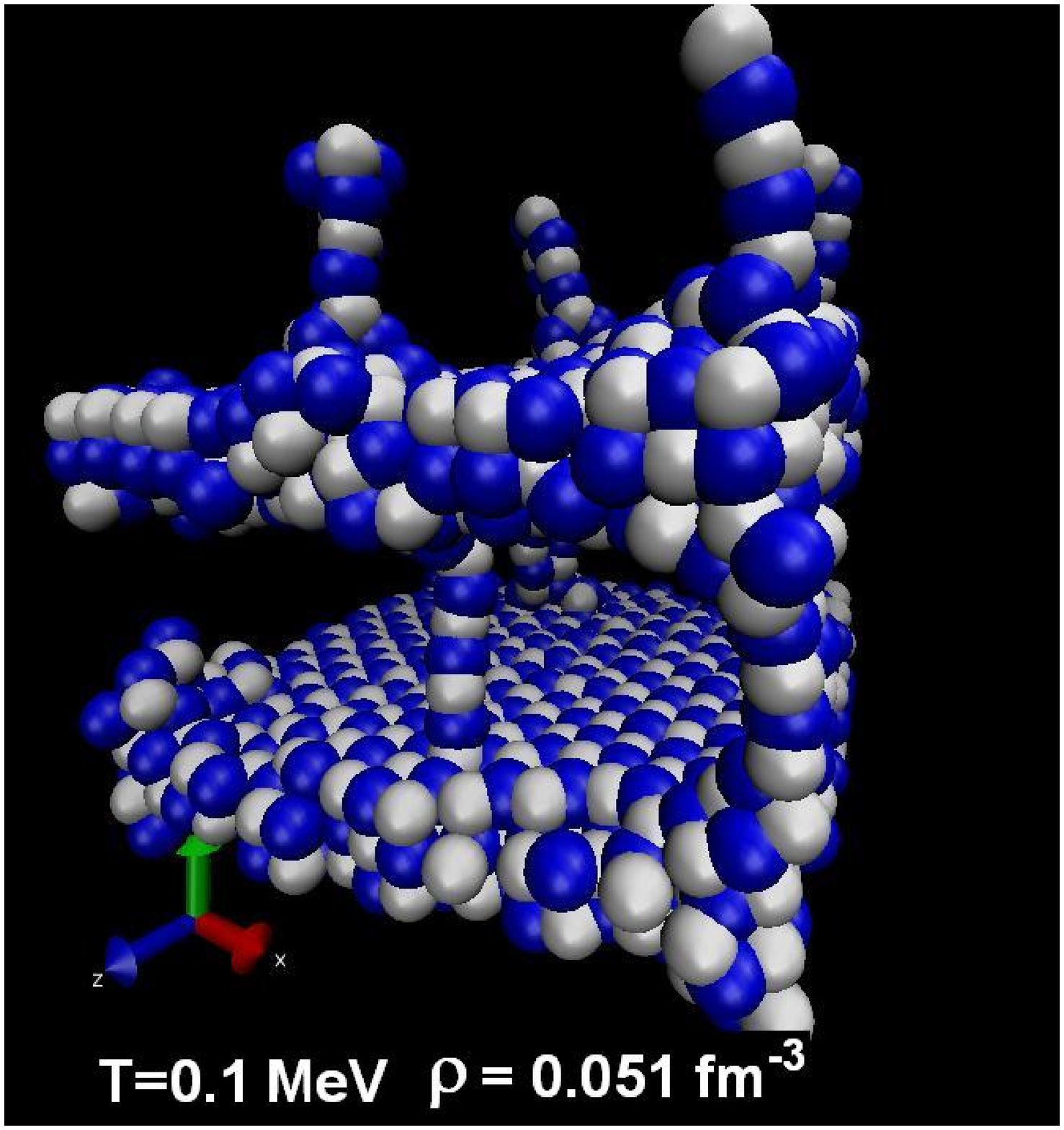}
\end{array}$
\end{center}
\caption{Configurations obtained with $CMD$ with $x=0.5$ with the
screened Coulomb potential for temperatures cooling from $T=1.0 \
MeV$, down to $T=0.1 \ MeV$.}\label{fig0}
\end{figure}

To illustrate both the effect temperature has on the nucleon
dynamics as well as the manner in which $CMD$ operates, it is
pedagogically useful to present the structure of nuclear systems
at varying temperatures.  Figure~\ref{fig0} shows the formation of
a final structure with $x=0.5$ and $\rho=0.55 \ fm^{-3}$ using the
screened Coulomb potential (to be presented in the next section)
as the system cools down from $T=1.0 \ MeV$ to $T=0.1 \ MeV$.

\subsection{Characterizing the structure of nuclear matter}
Once a configuration is achieved, the nucleon information is
recorded to provide the position and momentum of the nucleons
during the evolution.  Such microscopic information can be used to
identify clusters as well as to characterize the structure of
nuclear matter by means of the liquid structure function and the
Minkowski functionals.

\subsubsection{Cluster recognition}
The nucleon positions and momenta are used to identify the
fragment structure of the system by means of the "Minimum Spanning
Tree" ($MST$) cluster-detection algorithm of~\cite{20a} and
refined by~\cite{Str97}. In summary, $MST$ looks for correlations
in configuration space: a particle $i$ belongs to a cluster $C$ if
there is another particle $j$ that belongs to $C$ and $|r_i-r_j|
\leq r_{cl}$, where $r_{cl}$ is a clusterization radius which, for
the present study, was set to $r_{cl} = 3.0 \ fm$.

The main drawback of $MST$ is that, since only correlations in
$r$-space are used, it neglects completely the effect of momentum
giving incorrect information for dense systems and for highly
dynamical systems such as those formed in colliding nuclei.
Although more robust algorithms which look at relative momenta
between nucleons or pair-binding energies have been devised for
such systems (e.g. as the "Early Cluster Recognition Algorithm",
$ECRA$~\cite{dor-ran}), in the case of relatively cold systems,
such as nuclear crusts, the $MST$ is sufficient. In our case of
periodic boundary conditions, the $MST$ detection of fragments has
been modified to take into account the image cells and recognize
fragments that extend into adjacent cells.

\begin{figure}[h]
\begin{center}
\includegraphics[width=4.5in]{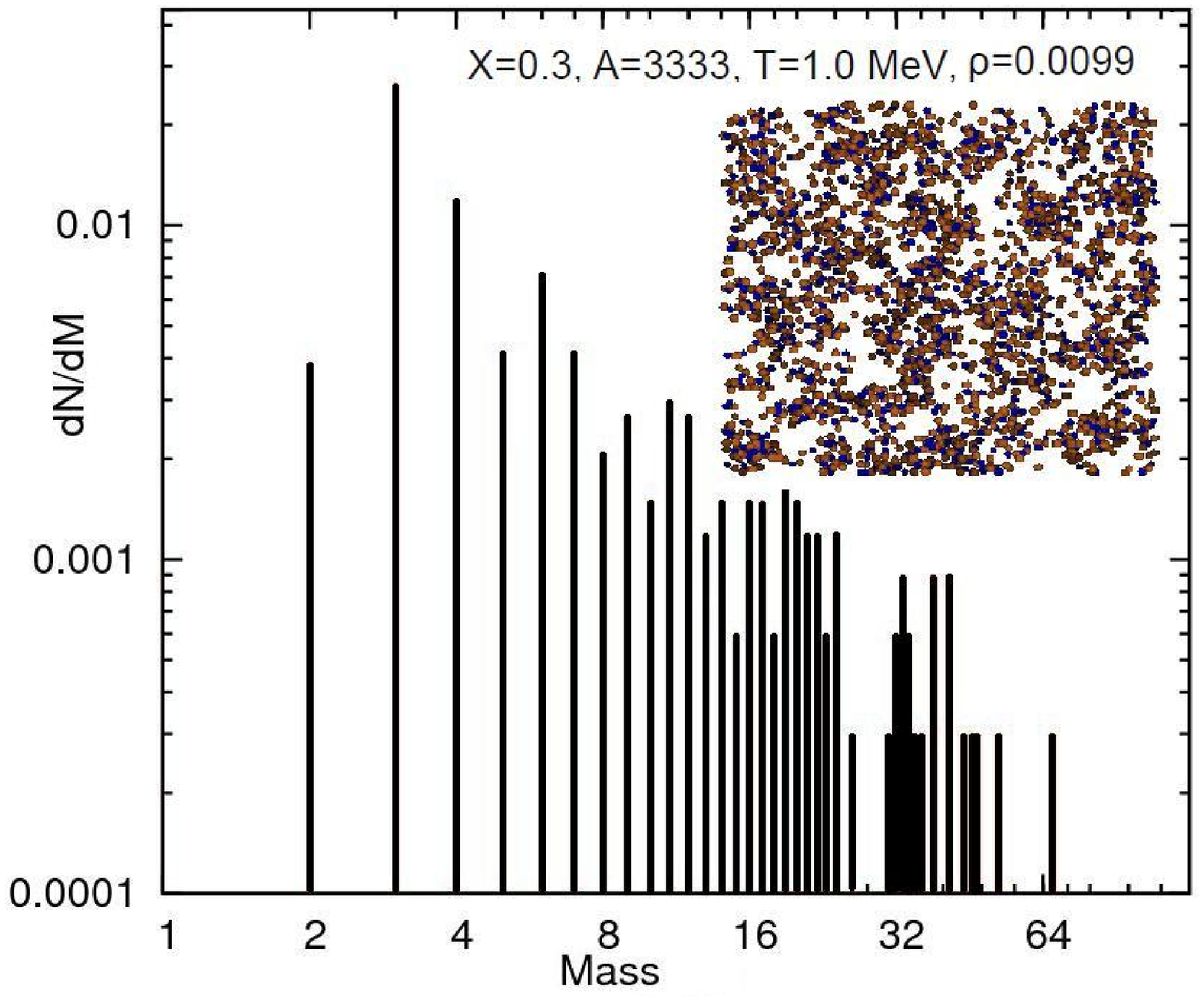}
\end{center}
\caption{typical size distribution of clusters as obtained with
$MST$.  The inset shows a projection of the particle spatial
distribution.}\label{histo}
\end{figure}

The Figure~\ref{histo} shows an example of the size distribution
of the clusters obtained for a case with $3,333$ nucleons,
$x=0.3$, and at $T=0.3 \ MeV$ and $\rho=0.009 \ fm^{-3}$.  The
inset shows a projection of the position of the nucleons within
the cell.  The shown structure was obtained with the ``screened
Coulomb'' treatment that will be described in the next section.

\subsubsection{Pair correlation function}
A further characterization of the structure of nuclear matter is
based on the liquid structure function, $S(k)$, which quantifies
the density fluctuations. $S(k)$ is obtained from the pair
correlation function, $g(r)$, which is the ratio of the average
local density to the global density, $g(r) = \rho(r)/\rho_0$.

The structure function is related to the Fourier spectrum of the
density fluctuations
\[
S(k) = (V/\rho_0)\left< \rho_k \rho_{-k}\right>,
\]
where $\rho_k$ denotes the Fourier transform of $\delta \rho(r)$.
This function can be obtained from the pair correlation function
$g(r)$ through
\[
S(k) = 1 + \rho \int d^3 r \exp \left(i {\bm k}\cdot {\bm
r}\right) \left[ g(r)-1 \right].
\]

For computing purposes, the pair correlation function $g(r)$ is
taken as the conditional probability density of finding a particle
at $\mathbf{r}_i+\mathbf{r}$ given that there is one particle at
$\mathbf{r}_i$. It gives information about the spatial ordering,
or structure, of a system of particles. Formally,
\[
g(r) = \frac{V}{4\pi r^2 N^2} \left\langle \sum_{i\neq j} \delta \left( r-r_{ij} \right) \right\rangle ,
\]
where $r_{ij}$ is $|r_i-r_j|$. For our case, this was calculated
by constructing histograms of the distances between particles for
several configurations and then averaging them. To keep
calculation times reasonable, but still attain accuracy at ranges
adequate to the study of the nuclear matter structure, we
considered particles (and their images) satisfying $r_{ij} \leq
1.5L$ with $L$ the simulation box size.

\begin{figure}[h]
\begin{center}
\includegraphics[width=4.5in]{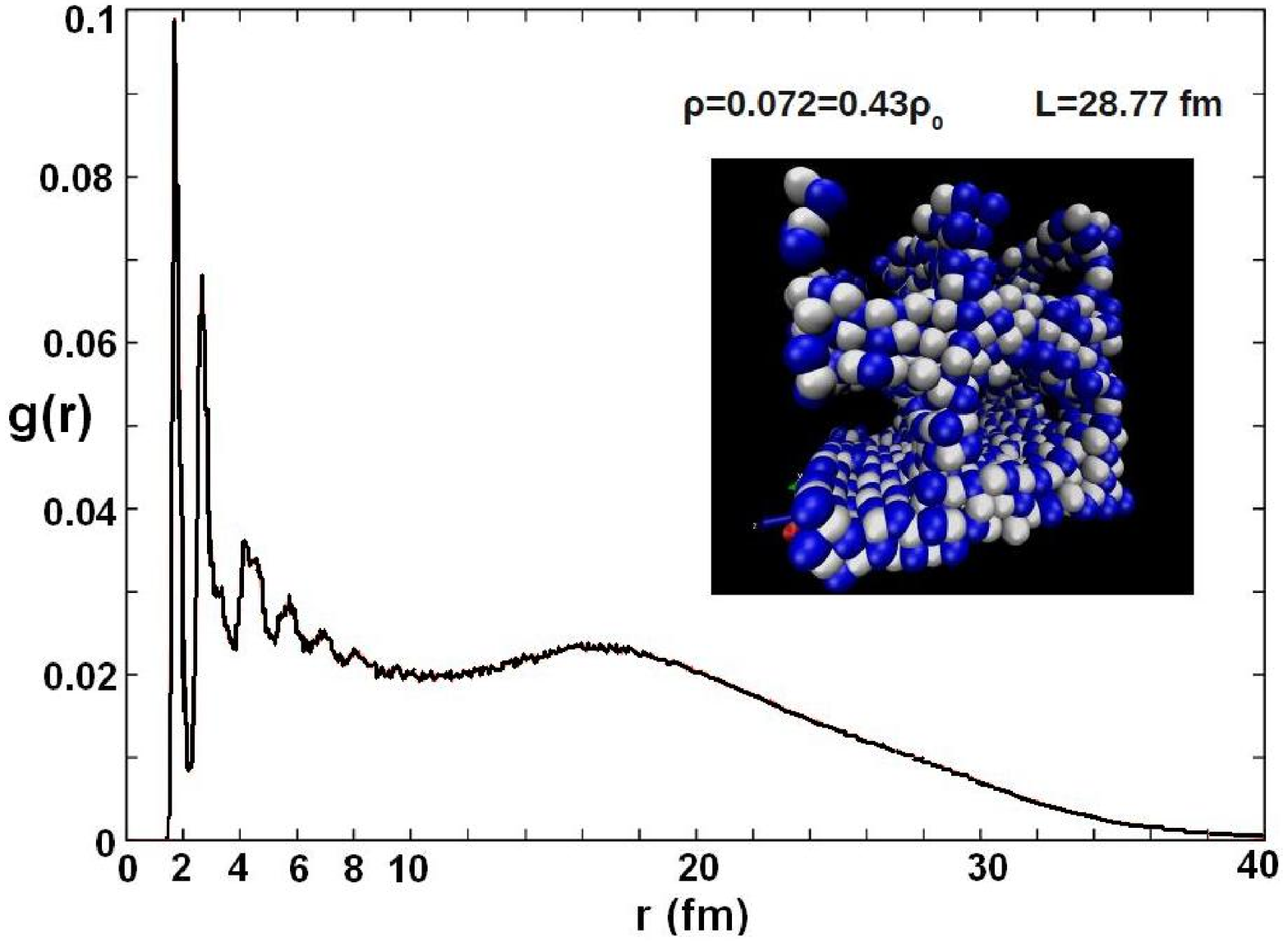}
\end{center}
\caption{Typical radial distribution function. The inset shows the
particle positions.}\label{g-r}
\end{figure}

The Figure~\ref{g-r} shows an example of the radial correlation
function obtained for a case with $2,000$ nucleons, $x=0.5$, and
at $T=0.1 \ MeV$ and $\rho\approx 0.043 \rho_0$.  The inset shows
three dimensional depiction of the position of the nucleons in the
$L=28.77 \ fm$ cell; again, this structure was obtained with the
``screened Coulomb'' treatment.

\subsubsection{Minkowski functionals}
Yet another method to characterize the morphology of
three-dimensional patterns in terms of geometrical and topological
descriptors are the Minkowski functionals.  Here a brief
practitioner's guide will be presented; for a full description of
the method, the reader is directed to~\cite{michielsen}.

The basic goal of the Minkowski functionals is to characterize the
topology of a distribution of points by means of quantifying
connected points as well as empty spaces; for this the patterns
are embedded in a uniform grid. For our case of a large number of
particles, the technique is applied by placing a grid inside the
$CMD$ cell and take the occupied grid cells as filled and those
not occupied as empty. Once this grid has been created the Euler
characteristic $\chi$ can be obtained as follows.

\begin{figure}[h]
\begin{center}
\includegraphics[width=4.5in]{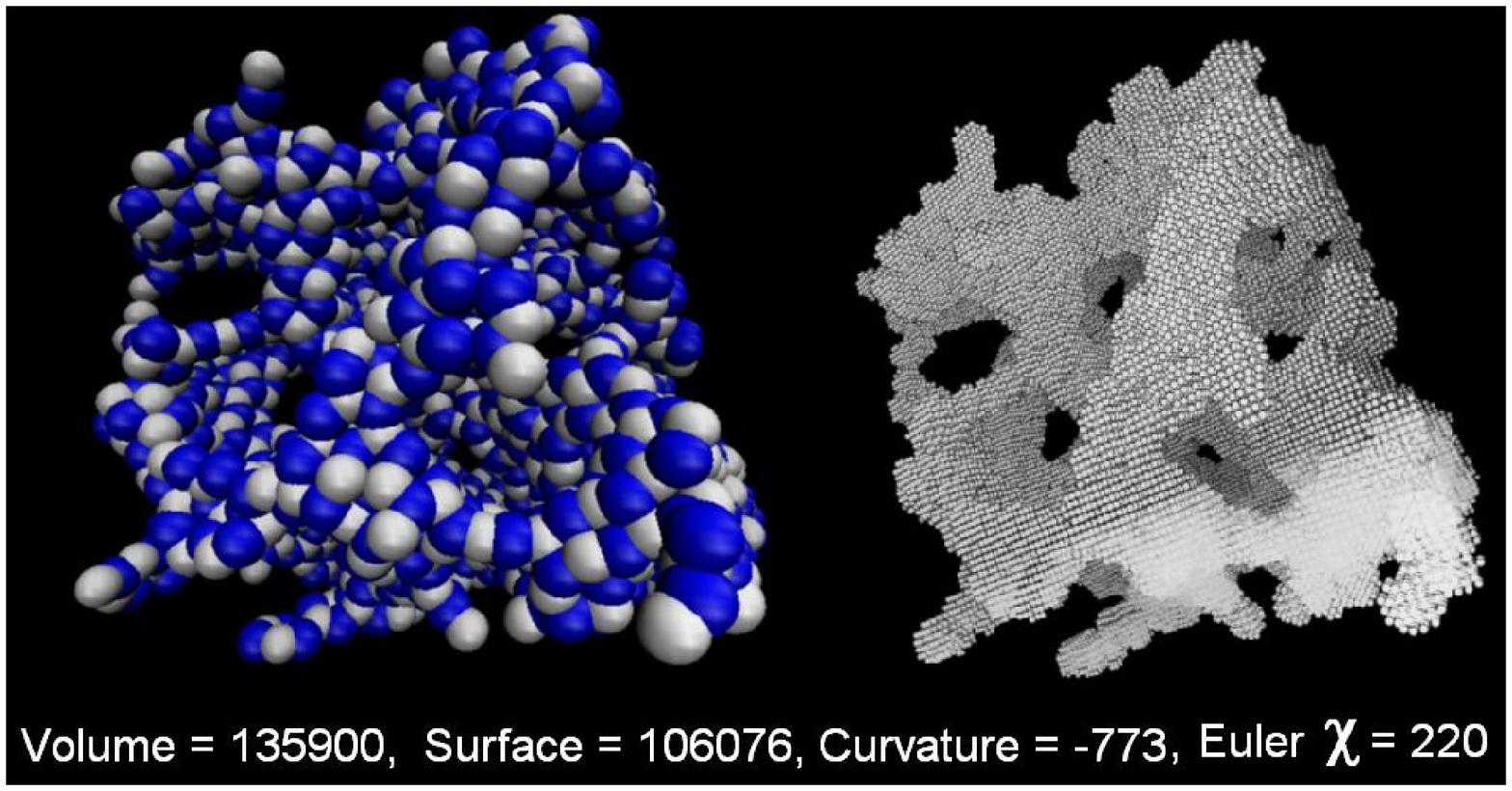}
\end{center}
\caption{Typical nuclear structure and the grid used to calculate
the Minkowski functionals.  The structure corresponds to a case
with $x=0.5$, $\rho=0.33 \ fm^{-3}$ and $T=0.1 \ MeV$.}\label{chi}
\end{figure}

In general, $\chi$ equals the number of regions of connected grid
cells minus the number of completely enclosed regions of empty
grid cells. Two grid cells are connected if they are immediate
neighbors, next-nearest neighbors, or are connected by a chain of
occupied grid cells. Characterizing the connected structure by its
number of occupied cubes, $n_c$, faces, $n_f$, and vertices,
$n_v$, including possible contributions from the interior of the
structure, the Minkowski functionals can be calculated through
\[
V=n_c, \ S = -6n_c+2n_f, \ 2B = 3n_c-2n_f+n_c, \
\chi=-n_c+n_f-n_c+n_v
\]
Figure~\ref{chi} shows an example of a nuclear structure, the
cubic grid enclosing it, and the values of the Minkowski
functionals.  For the computational aspects, the reader is
directed to~\cite{michielsen}.

\section{The Coulomb problem}\label{cp}
A major difference between the study of nucleon dynamics in
reactions and in stellar crusts is the presence of an embedding
electron sea. The nuclear system created in the outer part of a
neutron star is expected to be embedded in a relativistic
degenerate gas of electrons resulting from the large number of
weak decay processes that took place during the collapse-driven
supernova explosion.  The electron gas effectively renders the
system neutral and $\beta$-equilibrated.

The balance between nuclear and Coulomb interactions which
stabilize spherical nuclei is modified in the presence of the
electron gas.  The nuclear matter, be this in any type of shape or
form, now feels the additional Coulomb interactions of the
electron gas, which provides an effective screening and an overall
reduction of the Coulomb energy of felt by the protons.  The main
effect of these energy shifts is to modify the preferred nuclear
matter structure for a given density-temperature region; given the
infinite range of the Coulomb interaction some approximation is
needed when taking this effect into account in either static or
dynamical models; the effect has been studied, for instance, using
a $T=0$ liquid-drop model~\cite{gw-2003}, or with $QMD$ using a
screened Coulomb potential~\cite{7} or an Ewald summation
procedure~\cite{wata-2003}.

The fact that the results of these Coulomb calculations appear to
be model dependent~\cite{wata-2003} motivated us to present in
this review the two most common approaches, screened potential and
Ewald summation, using the same dynamical model, $CMD$. Concisely,
the $CMD$ model of protons and neutrons will now be assumed to be
immersed in a uniform gas of non-interacting electrons with the
exact electron density needed to guarantee charge neutrality. The
only effect of this sea of electrons is to add an overall Coulomb
interaction to the nuclear potential, and take into account its
effect in the equations of motion; for this we use two strategies,
Thomas-Fermi Screening and Ewald Summations, which are now
described in turn.

\subsection{Thomas-Fermi Screening}

The Thomas-Fermi Screening model of the Coulomb interaction takes
the electron gas as an ideal Fermi gas at the same number density
as the protons. This electron gas, being uniform, does not exert
any force on protons but becomes polarized in their presence
effectively ``screening'' the proton charge.

In this framework, the Coulomb interaction can be included in the
equations of motion for nucleons by means of the screened Coulomb
potential obtained from solving the Poisson equation. This
procedure yields a potential of the form:
\[V_C^{(Scr)}(r)=\frac{e^2}{r}\exp(-r/\lambda)\]
where $\lambda$ is a screening length which effectively turns the
Coulomb potential into a finite-range potential suitable for our
$MD$ calculations. The relativistic Thomas-Fermi screening length
is given by
\[\lambda=\frac{\pi^2}{2e}\left(k_F \sqrt{k_F^2+m_e^2}\right)^{-\frac{1}{2} }\]
where $m_e$ is the electron mass, the electron Fermi momentum is
given by $k_F=\left( 3\pi^2\rho_e\right)^{1/3}$, and $\rho_e$ is
the electron gas number density (taken equal to that of the
protons). To avoid finite size effects, $\lambda$ should be
significantly smaller than the size of the simulation cell,
$L=\left( A/\rho\right)^{ 1/3 }$, but since $\lambda$ depends on
the density of the system, it is always possible to satisfy this
condition by increasing the simulation box size along with the
number of particles. However, since this can lead to prohibitively
large systems for our current computation capabilities, we follow
the prescription given in~\cite{horo_lambda} and set $\lambda=10 \
fm$.

\begin{figure}[h]
\begin{center}$
\begin{array}{cc}
\includegraphics[width=2.5in]{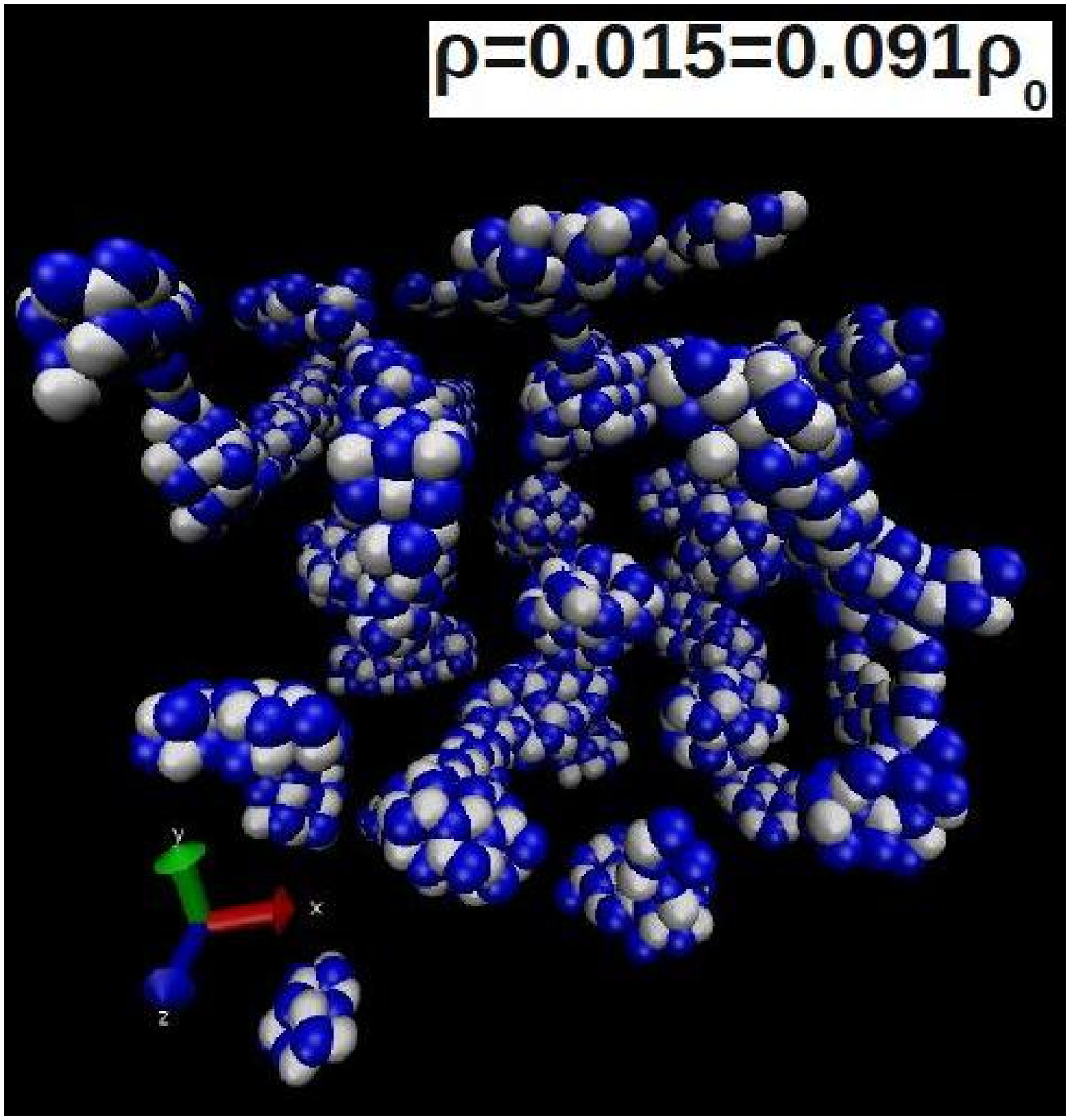} &
\includegraphics[width=2.5in]{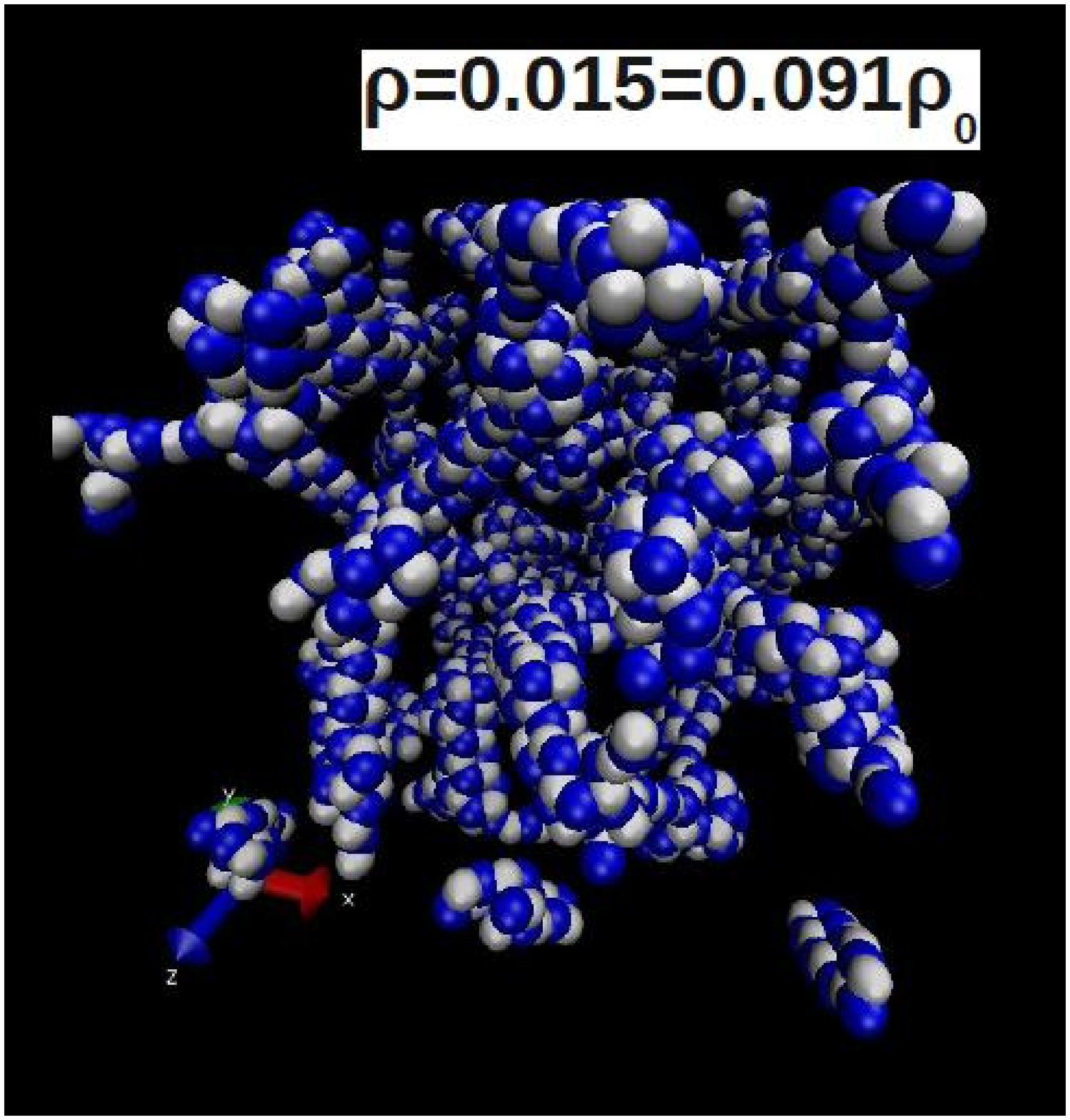} \\ \includegraphics[width=2.5in]{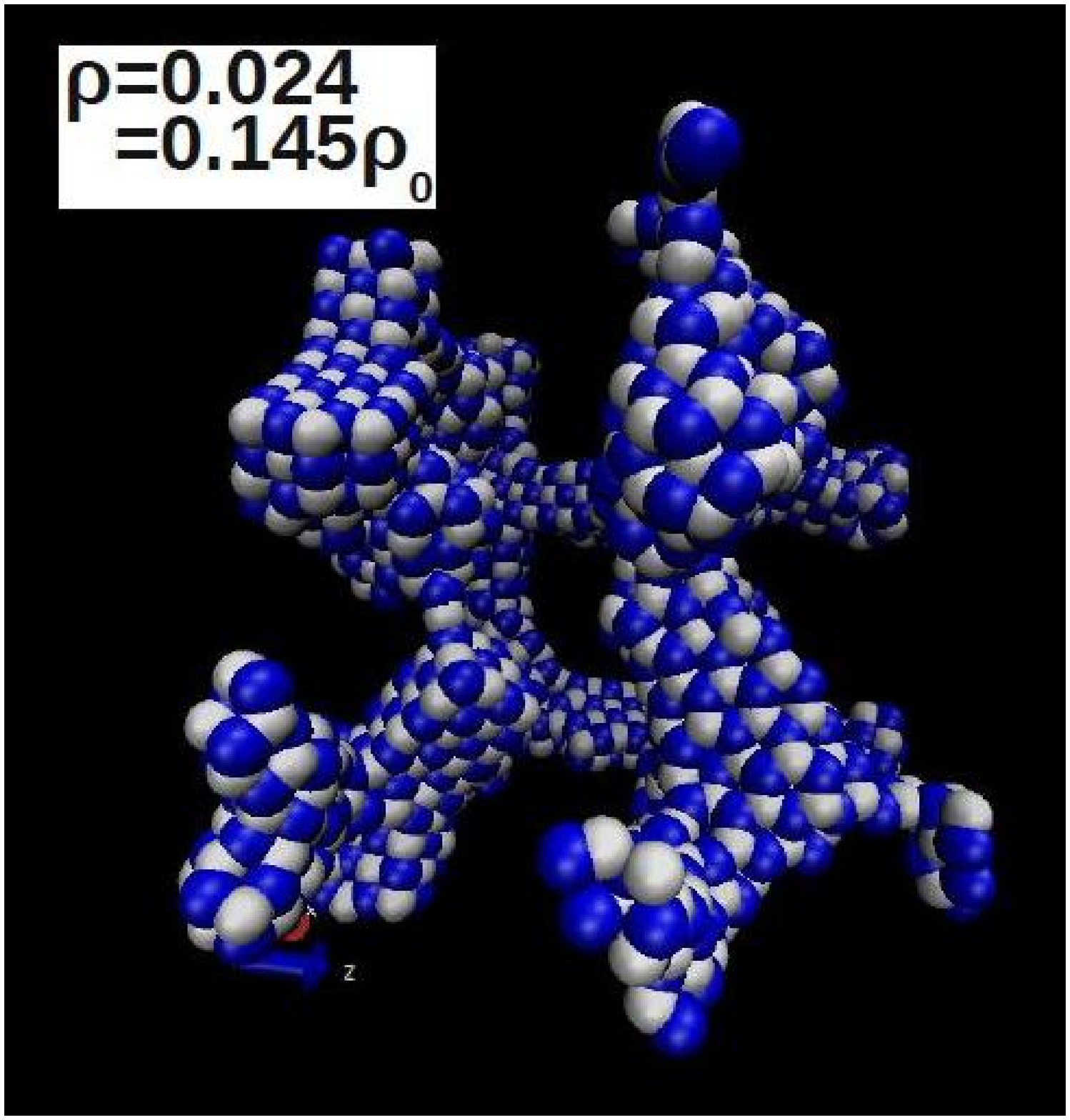} &
\includegraphics[width=2.5in]{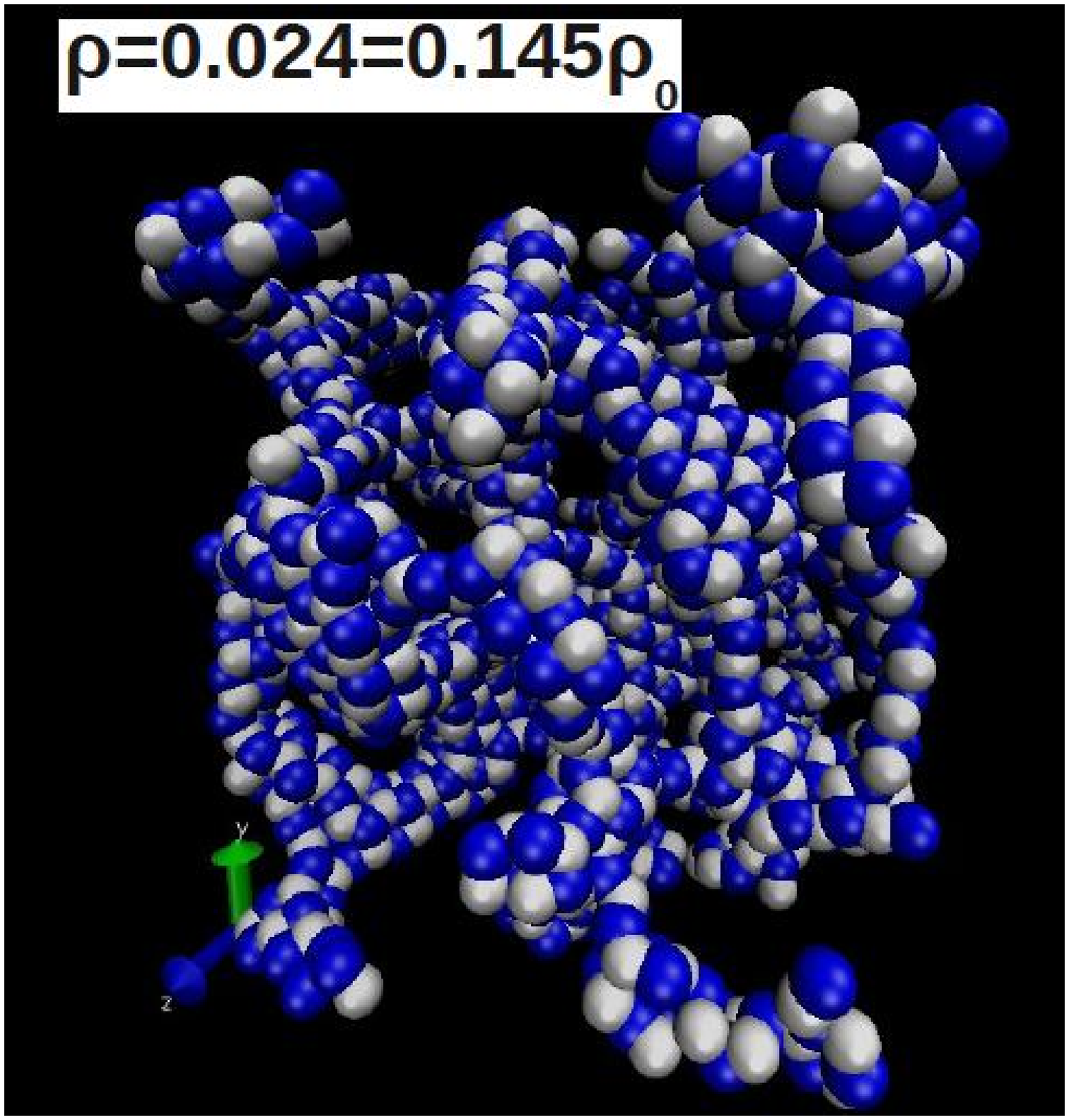} \\ \includegraphics[width=2.5in]{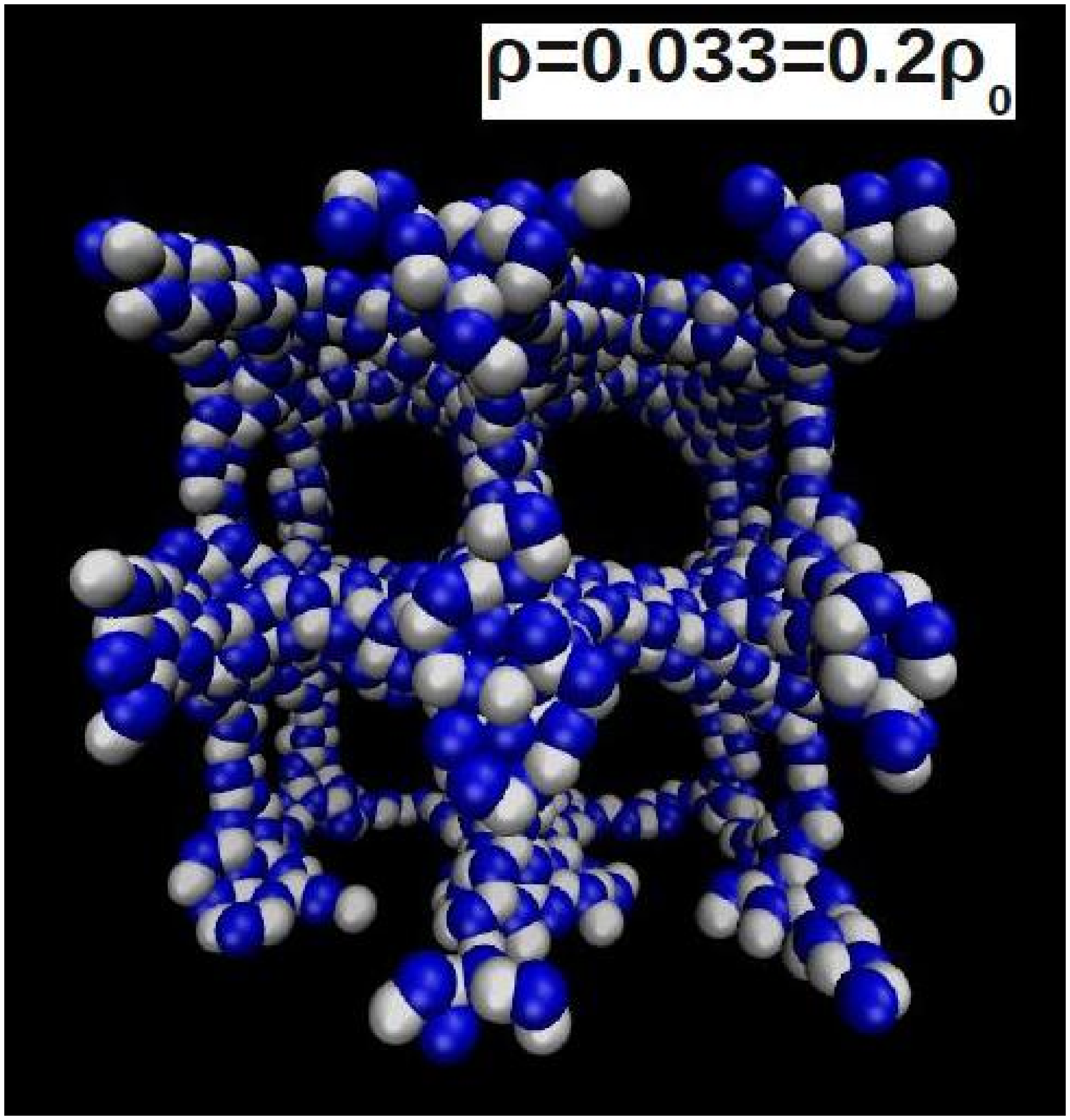} &
\includegraphics[width=2.5in]{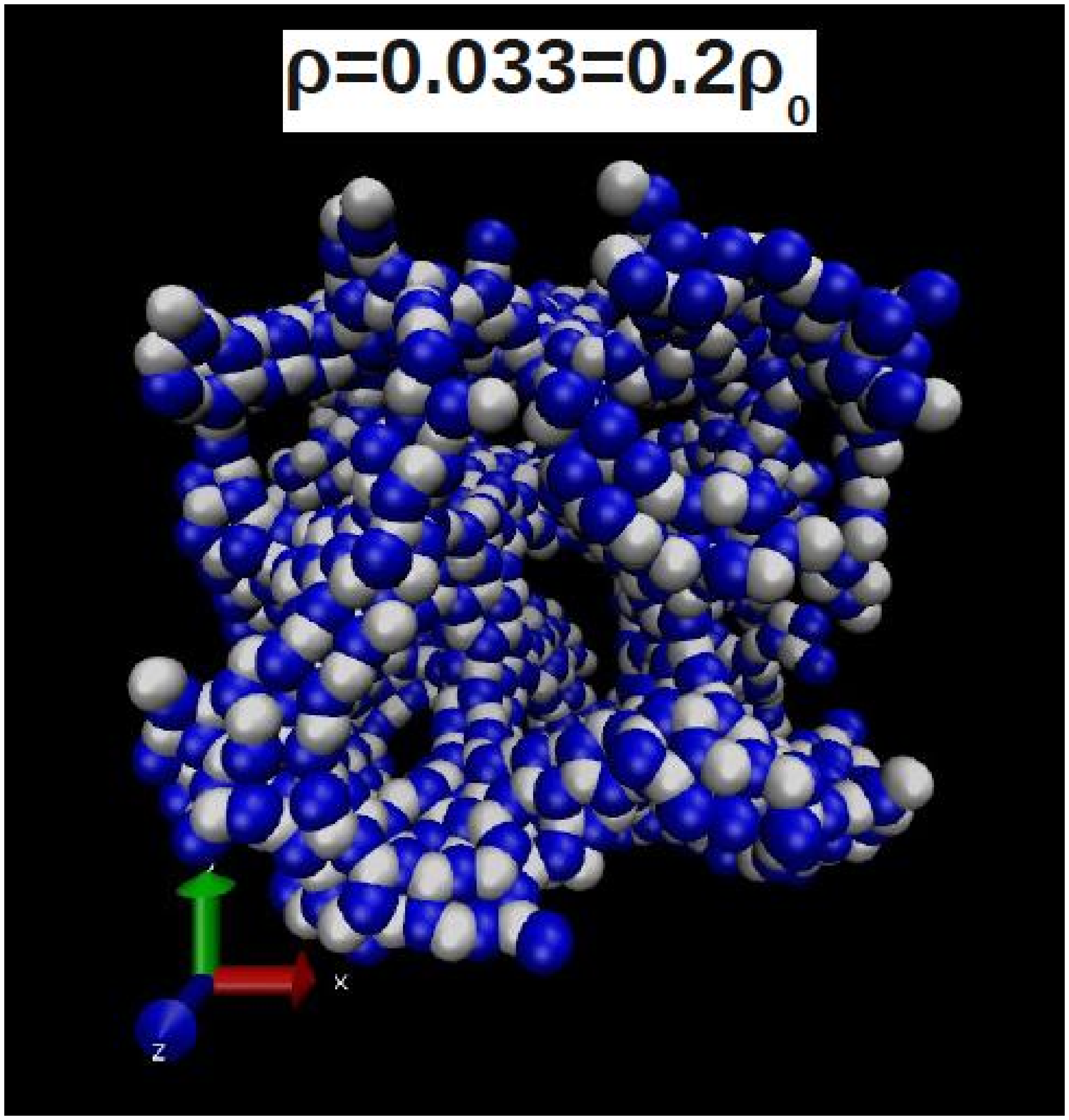}
\end{array}$
\end{center}
\caption{Configuration obtained with $CMD$ with $x=0.5$, $T=0.1 \
MeV$, with the screened Coulomb potential (left figures), and with
the Ewald summation (right figures).}\label{fig1}
\end{figure}

\begin{figure}[h]
\begin{center}$
\begin{array}{cc}
\includegraphics[width=2.5in]{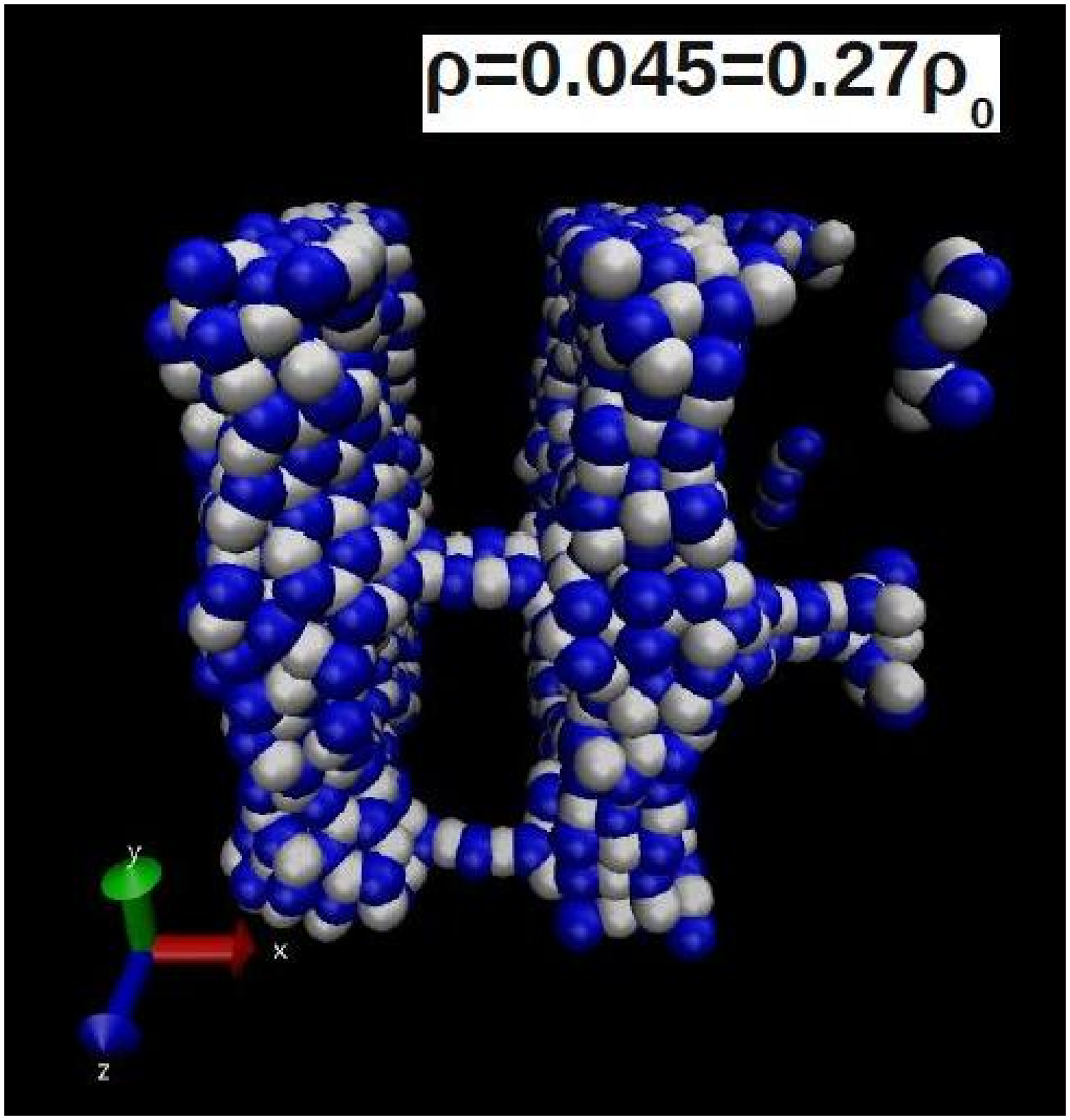} &
\includegraphics[width=2.5in]{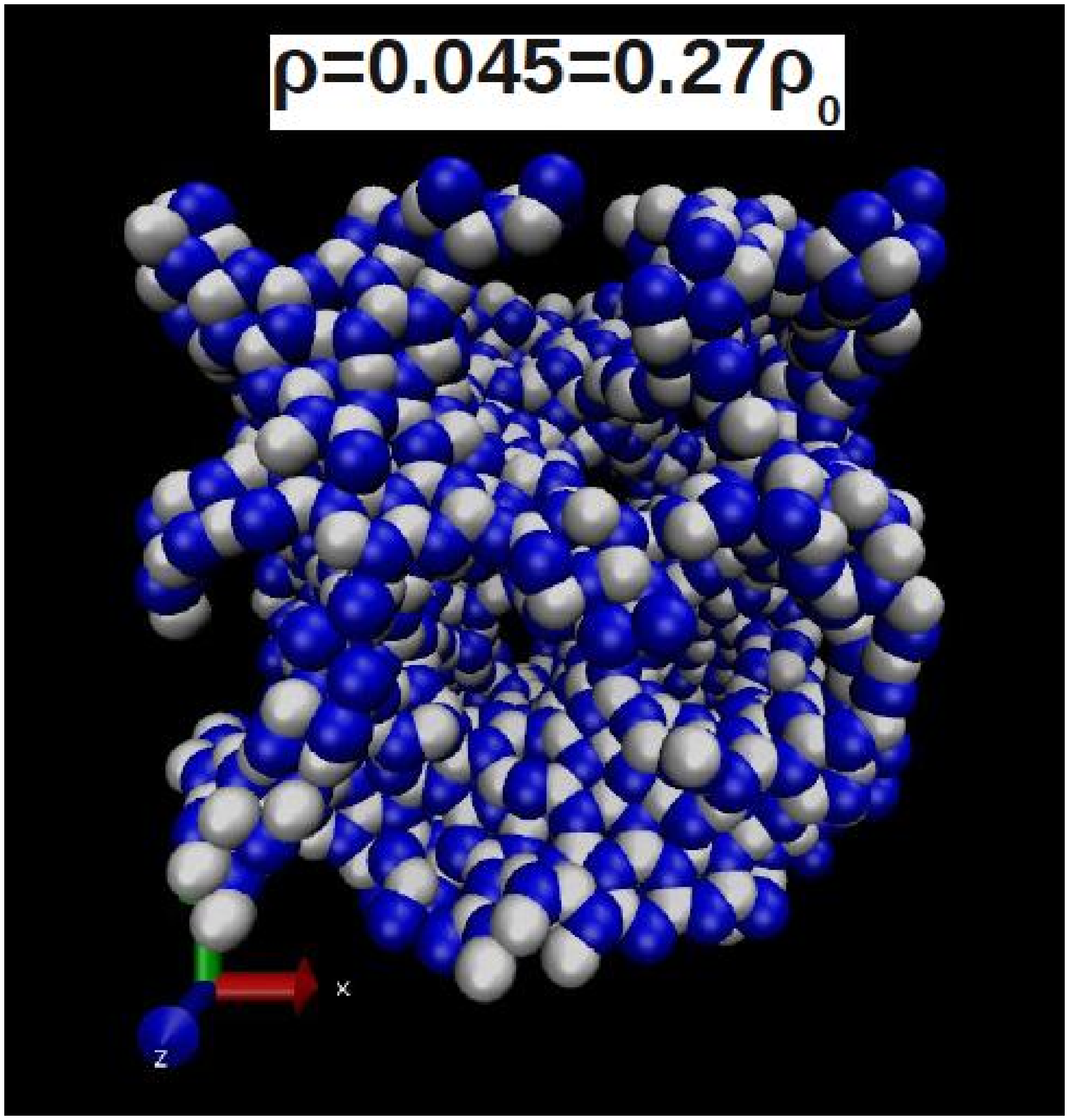} \\ \includegraphics[width=2.5in]{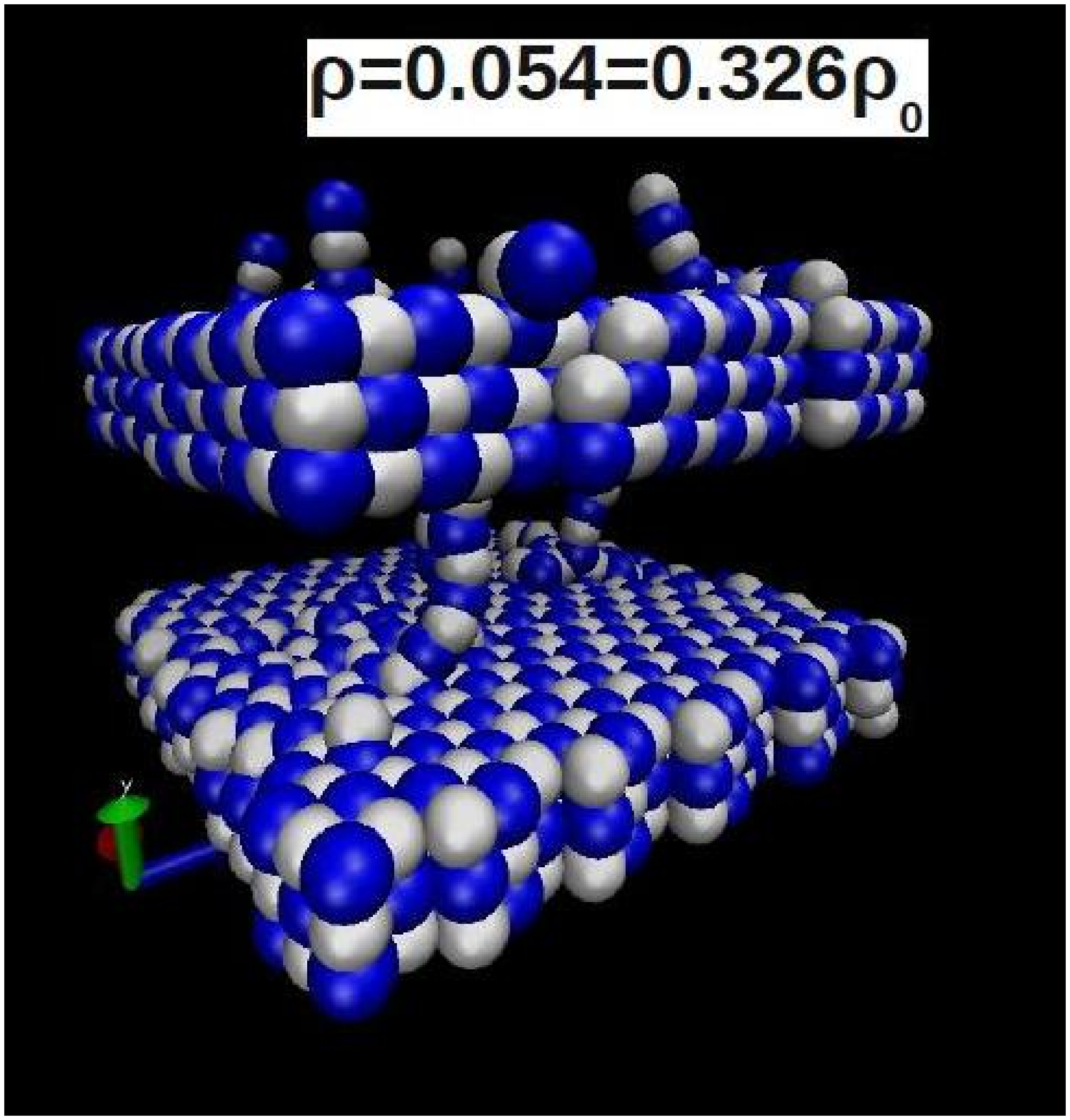} &
\includegraphics[width=2.5in]{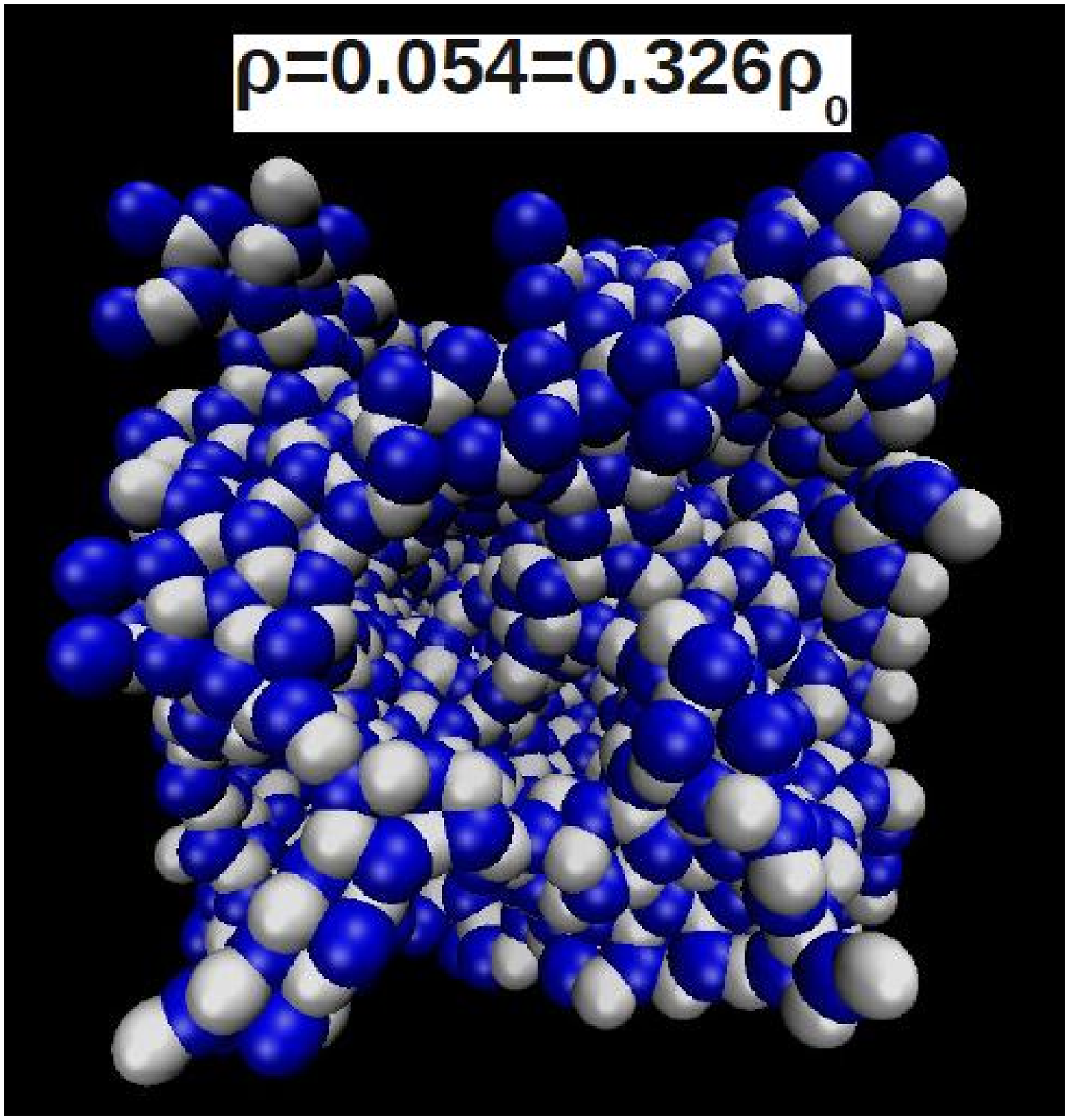} \\ \includegraphics[width=2.5in]{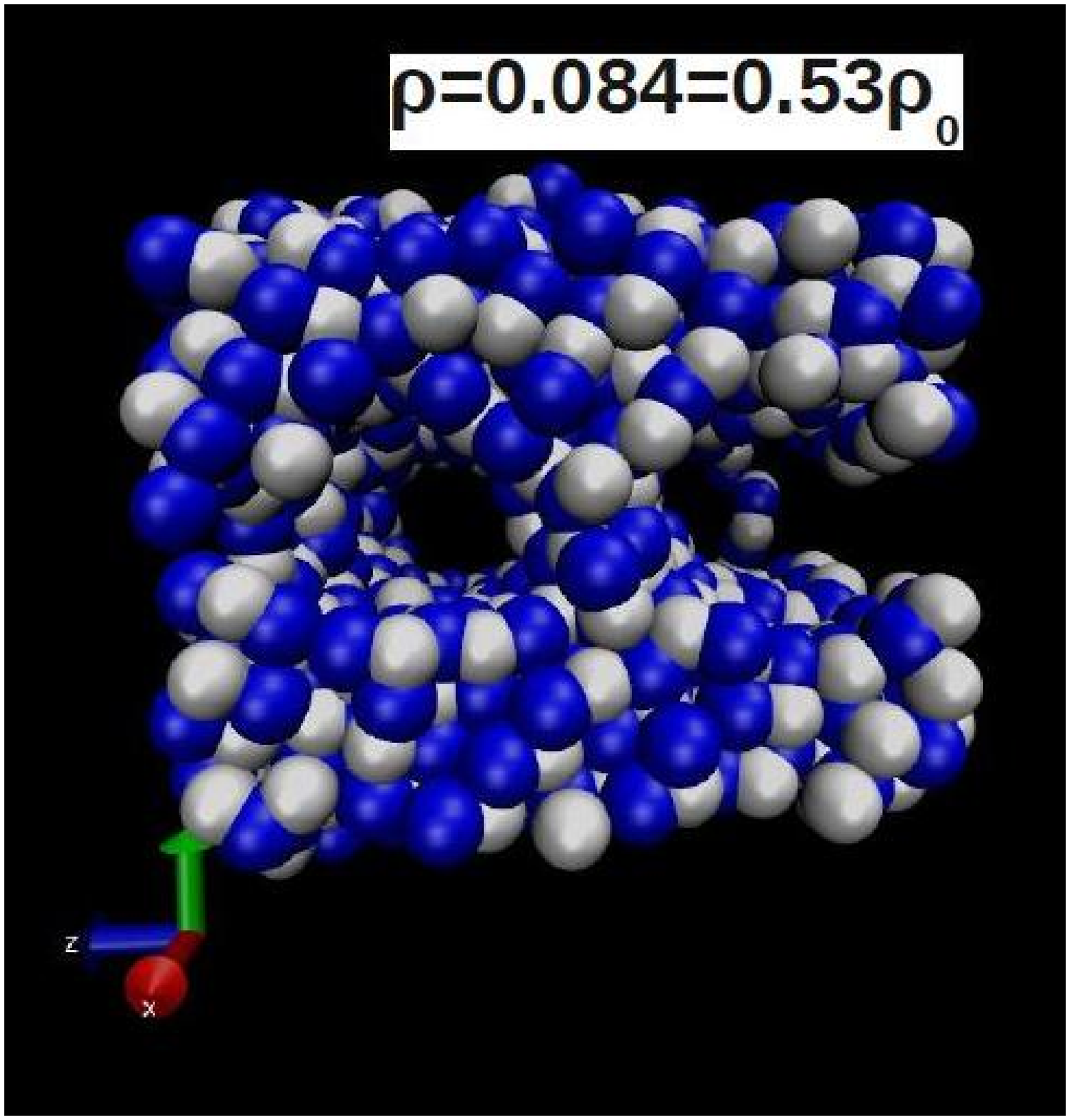} &
\includegraphics[width=2.5in]{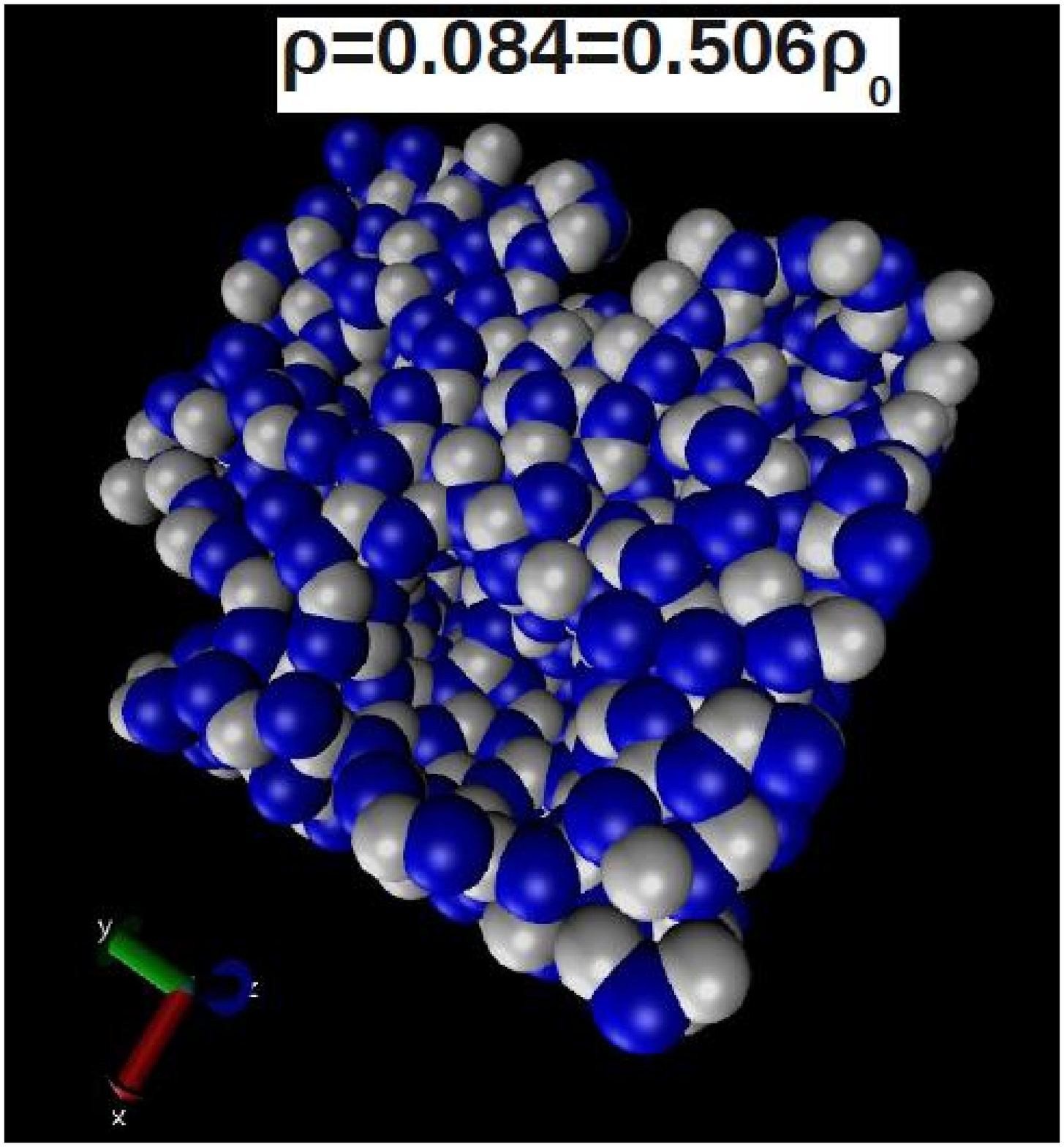}
\end{array}$
\end{center}
\caption{Configuration obtained with $CMD$ with $x=0.5$, $T=0.1 \
MeV$, with the screened Coulomb potential (left figures), and with
the Ewald summation (right figures).}\label{fig2}
\end{figure}

To compare to the results obtained with the Ewald summation to be
presented in the following section, figures~\ref{fig1} and
\ref{fig2} present on the left column representative structures
obtained with symmetric nuclear matter ($x=0.5$) using the
screened Coulomb method. For clarity, the figures exclude free
particles and show only the connected structures with four
nucleons or more.

\subsection{Ewald Summation}

The Ewald summation technique is widely used to compute long range
interactions; in this case, the background electron gas plays a
more subtle but equally essential role. In this approach, each
proton is surrounded by a Gaussian charge distribution of equal
magnitude and opposite sign that screens the original. Each of
these Gaussians is likewise cancelled by opposing distributions.
This allows for the problem to be split in two: a short-ranged
interaction between screened charges, and an infinite periodic
system of Gaussian charges that can be solved in reciprocal space
using Fourier transforms. We refer the reader
to~\cite{frenkel,nymand_linse} for details on the derivation.

The screening Gaussian distributions exactly compensate the proton
charge and have a width set by a parameter $\alpha$:
\[ \rho_{Gauss}(r) = -e \left( \frac{\alpha}{\pi}\right)^{\frac{3}{2}} \exp(-\alpha r^2) \]
where $e$ is the charge of a single proton. The real space part of
the contribution to the Coulomb potential energy takes the form
\[U_{Ewald}^{Real} = \frac{e^2}{2} \sum_{i\neq j}^Z \frac{\textbf{erfc}(\sqrt{\alpha}r_{ij}) }{r_{ij}} - \left( \frac{\alpha}{\pi} \right)^\frac{1}{2} \sum_{i=1}^Z e^2 \]
where $Z$ is the number of protons and $\textbf{erfc}(x)$ the
complementary error function. The last (constant) term is a
correction for a ``self interaction'' between each proton and its
Gaussian partner. The parameter $\alpha$ controls the range of the
short range effective potential of the screened proton.

The contribution to the energy of the remaining charge
distributions is given by
\begin{equation}\label{eq_ewald_fourier}
 U_{Ewald}^{Fourier} = \frac{1}{2V} \sum_{\textbf{k} \neq 0}^{k_{co}} \frac{4\pi}{k^2} \left| \rho(\textbf{k}) \right|^2 \exp\left( -\frac{k^2}{4\alpha}\right)
\end{equation}
where
\[ \rho(\textbf{k}) = \sum_{j = 1}^Z \exp \left( i\textbf{k} \cdot \textbf{r}_j \right) \]
is the structure factor.

The summation in Eq.~(\ref{eq_ewald_fourier}) runs over all
reciprocal space vectors ($\textbf{k} = \frac{2\pi}{L} \left(n_x,
n_y, n_z\right)$ with $n_i$ integers) inside a sphere defined by
the cutoff $k_{co}$. The problematic term with $k=0$ corresponds
to a polarization energy that banishes because of infinite
polarizability of the background electron gas.

The Fourier space part is by far the most time consuming part of
the simulation. The choice of $\alpha$ and $k_{co}$ is crucial to
balance the length and the accuracy of a simulation. We chose
$\alpha$ so that the range of the real space Ewald sum matches
that of the nuclear part and $k_{co}=\frac{2\pi}{L} n_c$ with
$n_c=8$. This guarantees a relative error in the energy of
$\mathcal{O}(10^{-6})$.

Figures~\ref{fig1} and~\ref{fig2} show representative structures
obtained with symmetric nuclear matter ($x=0.5$) using the Ewald
summation method with systems at densities varying between
$0.09\rho_0 < \rho < 0.5\rho_0$. As in the previous figures, free
particles are not included and only connected structures with four
nucleons or more are shown.

\subsection{Discussion}
The differences of the structures obtained with the screened
Coulomb method and the Ewald summation technique are readily
noticeable.  The screened potential allowed the formation of the
whole menu of pasta types, from polpetta at low densities, to
spaghetti, to lasagna, gnocci and to scaciatta (Italian meat pie,
i.e. a uniform state with holes) at higher densities.

Comparing to previous results (at lower values of $x$) we observe
a similar evolution of the structures bt at different values of
the density. Ref.~\cite{4}, for instance, produces uniform states
with holes (scaciatta) at $0.6\rho_0-0.7\rho_0$ while here such
structure is observed at $0.4\rho_0-0.5\rho_0$, we find the
lasagna-type layers between $0.2\rho_0-0.3\rho_0$ while studies
using Wigner-Seitz~\cite{4} cells place such a structure at
$0.4\rho_0-0.5\rho_0$ and even at $0.105\rho_0$~\cite{wata-2003},
likewise we observe the spaghetti and meatballs phases at
$0.05\rho_0-0.15\rho_0$ while other studies~\cite{wata-2003}
obtain such structures at $0.06\rho_0-0.09\rho_0$.

The structures obtained with the Ewald summation presented a much
reduced menu.  Spaghetti rods appeared briefly at very low
densities ($0.03\rho_0 - 0.1\rho_0$), sp\"atzle-like or
gnocci-like structure at intermediate densities ($0.15\rho_0 -
0.2\rho_0$), and scaciatta at all other higher densities
($>0.3\rho_0$).

The differences between the results obtained with different
Coulomb approximations evidence the need to use a proper method to
evaluate the Coulomb interaction of the electron gas. The
extension of this study to other proton ratios and other
temperatures is in progress; preliminary results indicate that
neutron-rich matter (with $x=0.3$) has a reduced range of
possibilities, mostly limited to lasagna-like and gnocci-like
structures. These and other results of the present study (pair
correlation function, fragment distribution, free neutrons,
Minkowski functionals, etc.) will be published elsewhere in the
near future.

\section{Final remarks}\label{fr}
Neutron stars have a crust composed of neutron-rich nuclear matter
immersed in a sea of electrons.  At subnormal densities, the
structure of such crust is determined by a fine balance of nuclear
and electric forces, and it goes from a uniform consistency at
normal densities, to a ``clumpy'' texture, to rod-like, and to
layer-like as the density decreases.  Such structures have been
studied with static and dynamic models, and in this article the
classical molecular dynamics model is used to illustrate several
features of this ``nuclear pasta''.

The $CMD$ has been extremely successful to reproduce many features
of cold nuclei, nucleon scattering, as well as heavy ion
collisions; its extension to infinite systems allows its
application to stellar crusts.  The study of the pasta phases
requires the use of cluster recognition algorithms, radial
correlation functions and other topological instruments, such as
the Minkowski functionals.  Other required techniques are tools to
introduce the effect of the electron sea into the $CMD$ equations
of motion, we reviewed two such methods: the screened Coulomb
potential and the Ewald summation and found substantial
differences in the structures formed under the two methods: those
formed with the screened potential ARE much richer in variety than
those obtained with the Ewald summation.

An implicit result is the fact that the $CMD$ is a computationally
efficient method to study the nuclear pasta.

\section*{Acknowledgments}
C.O.D. is a member of the "Carrera del Investigador" CONICET
supported by the Universidad de Buenos Aires, CONICET through
grant PIP5969.  J.A.L Acknowledges support from grant NSF-PHY
1066031.  J.A.L. thanks Dr. Jorge Piekarewicz for suggesting the
use of $CMD$ to study the nuclear pasta.

\label{lastpage-01}

\end{document}